\documentclass[aps,pre,reprint]{revtex4-1}

\usepackage{amsmath}
\usepackage{amssymb}
\usepackage{graphicx}
\usepackage{color}
\usepackage{bm}
\usepackage{dcolumn}

\newcommand{\kb} {k_{\scriptscriptstyle B}}

\begin{document}

\title{Classical molecular dynamic simulations and modelling of\\ inverse-bremsstrahlung heating in low Z weakly-coupled plasmas}
\author{\surname{R.} Devriendt}
\affiliation{CEA, DAM, DIF, F-91297 Arpajon, France}
\author{\surname{O.} Poujade}
\affiliation{CEA, DAM, DIF, F-91297 Arpajon, France}
\affiliation{Université Paris-Saclay, CEA, LMCE, F-91128, Bruyères-le-Châtel, France}

\date{\today}

\begin{abstract}

Classical molecular-dynamics simulations (CMDS) have been conducted to investigate one of the main mechanism responsible for absorption of radiation by matter namely stimulated inverse bremsstrahlung. CMDS of two components plasmas (electrons and ions) for a large range of electron densities, electron temperatures, for ionization $Z=1$, were carried out with 2 million particles using the code LAMMPS. A parameterized model (with 6 adjustable constants), which encompasses most theoretical models proposed in the past to quantify heating rate by stimulated inverse bremsstrahlung, serves as a reference for comparison to our simulations. CMDS results are precise enough to rule out elements of these past models such as coulomb logarithms depending solely upon laser pulsation $\omega$ and not upon intensity. The 6 constants of the parameterized model have been adjusted and the resulting model matches all our CMDS results and those of previous CMDS in the literature.

\end{abstract}

\maketitle 

\section{Introduction}

Inverse bremsstrahlung (IB) is the process of absorption of a single photon by a free electron in the field of another particle (ion or neutral atom). It is the main source of absorption of laser light by matter for intensities less than \mbox{$10^{16}$ W/cm$^2$}. It is far from being the only effect responsible for laser absorption directly or indirectly. Non resonant ponderomotive effects (laser beam self-focusing, filamentation) or resonant ponderomotive effects (Brillouin, Raman, parametric decay and oscillating two-stream instabilities, two plasmon decay, Langmuir cascade, two phonon decay of phonon, etc) also contribute to absorption in their own way but IB is the most important. 

\par 

This process of IB heating is modeled in several different ways in the extensive literature on the subject. Theoretical works are either based upon a classical approach \cite{landau1936, daw62,silin65,PF-Johnston-1973,jon82,skup87, mulser01, brant03} or upon a quantum approach \cite{rand64,shim75,scle79,sil81,skup87,meyer94,kull01,brant03, moll14}. In both cases, there does not seem to be a general agreement, and quantum model do not converge to the classical limit for vanishingly small values of $\hbar$. Few studies, such as \cite{bunk73, seel73, brant03}, even go so far as to take a critical look at some of them. 

\par 

In order to challenge these theoretical results, numerical evaluations of IB heating has been carried out but mostly using Fokker-Planck (FP) simulations \cite{matte84, ersf00, weng06, weng2009, le19}. These simulations require collision kernels to be specified which amounts to making assumptions at microscopic level.

\par 

Ab initio classical molecular dynamic simulation (CMDS) of IB are very few in the recent literature and this is one of the objective of this article, to strengthen the share of these CMDS with present high-performance-computing capabilities. The second objective is to provide a literal expression for the IB heating $\mathrm{d}(\kb T_e)/\mathrm{d}t$ and, as a corollary, for the electron-ion collision frequency in the IB process $\nu_{ei}^{\mathrm{IB}}$ (there is no reason it should be the same as the electron-ion collision frequency within other mechanisms such as temperature relaxation $\nu_{ei}^{\mathrm{T}}$ \cite{Dimonte2008} or velocity relaxation $\nu_{ei}^{\mathrm{V}}$ \cite{baal19}, both of which will be described hereafter). 

\par 

There are several different such expressions in the literature and we are now at a point where it is possible to reliably discriminate these expressions by use of microscopic molecular dynamic simulations. Such a literal expression is of the essence when it comes to simulate the interaction of intense radiations with matter within complex flows using radiation-hydrodynamics codes \cite{troll2018, POP-Marinak-2001, lasnex}, particle-in-cell codes \cite{smilei18, calder03} or Fokker-Planck codes.  

\par 

Most theoretical expressions of inverse-bremsstrahlung heating in the literature can be summarized in a single parameterized expression that will be presented in \S\ref{sec:param}. Numerical simulations dedicated to IB absorption in the literature will be presented in \S\ref{sec:num}. The classical modeling of a two component plasma described in our simulations and the description of our CMDS will be the subject of \S\ref{sec:mod} and \S\ref{sec:lammps}. Finally, results of our CMDS on situations without oscillating electric field will be compared to existing results \cite{Dimonte2008, baal19} to ascertain our simulation settings in \S\ref{sec:dim} and the determination of the adjustable constants of the parameterized model will be carried out in \S\ref{sec:fin}.  

\section{Theoretical modeling for inverse bremsstrahlung absorption in the literature}\label{sec:param}

Classical solutions to this problem have been provided through different techniques in the literature. The oldest was by way of ballistic modeling as in Landau \cite{landau1936} or Mulser \cite{mulser01} where one considers the trajectory of one single electron in the field of a single screened scattering center. The second technique makes use of Vlasov equation, for irradiation frequencies near the plasma frequency, as in the work of Dawson, Oberman and Johnston \cite{daw62, daw64, PF-Johnston-1973}. The last techniques consists in solving the Boltzmann equation with a Lenard-Balescu collision term in order to get a solution for a wider range of frequencies and intensities as in Silin's work \cite{silin65}. This last method was also used by Jones and Lee \cite{jon82} to discuss the evolution of the electron velocity distribution in a plasma heated by laser radiation. In a nutshell, these theoretical works all led to similar template formulas of the electron heating rate by inverse Bremsstrahlung 
\begin{align}
\frac{\mathrm{d} (\kb\,T_e)}{\mathrm{d}t}=\frac{2}{3 n_e}\frac{\omega_p^2}{c\,\omega^2}\,\nu_{ei}^\mathrm{IB} \,I\label{eq:heatrate}
\end{align} where $I$ and $\omega$ are respectively the radiation intensity and pulsation and $\nu_{ei}^\mathrm{IB}$ is the electron-ion collision frequency for the IB process that is proportional to a Coulomb logarithm, $\ln(\Lambda_\mathrm{ei}^\mathrm{IB})$ as can be seen in the general formula (\ref{eqnuei}). 

\par 

In this classical context, there is, of course, no dependence upon $\hbar$. All these analytical expressions have been derived assuming that electrons velocity distributions are Maxwellians. Collective effects are hidden in the Coulomb logarithm which is generically of the form $\ln(\Lambda)=\ln(b_\mathrm{max}/b_\mathrm{min})$ where, in the absence of irradiation ($I=0$), $b_\mathrm{max}$ corresponds to the range of collective interactions (of order the Debye length \mbox{$\lambda_D=\sqrt{\varepsilon_0\kb T_e/e^2 n_e}$}) and $b_\mathrm{min}$ corresponds to the shortest distance accessible to these charged particles (of order either the closest distance of approach, also known as the Landau length \mbox{$R=Z\,e^2/4\pi\varepsilon_0\,\kb T_e$}, or the de Broglie length). The fuzziness of the expression of $\ln(\Lambda)$ is characteristic of theoretical calculations or reasonings where collective effects of all particles on all other is not properly addressed from first principles and relies on the assumption that it can, in a sense, be captured by studying the motion of one electron around one ion screened by the mean field of all other charged particles (all other electrons and ions) which is largely disputable. 

\par

The Coulomb logarithm in the absence of laser irradiation is well known for the temperature relaxation and velocity relaxation. It has been derived from first principles using dimensional regularization \cite{bps05} and was confirmed to a very good accuracy by CMDS \cite{Dimonte2008, baal19}. On the contrary, for a plasmas submitted to laser irradiation, none of the theoretical studies listed in previous sections \cite{landau1936, daw62,silin65,PF-Johnston-1973,jon82,skup87, mulser01, brant03} give any precise formulation of $\ln(\Lambda)$ apart from the generic $\ln(b_\mathrm{max}/b_\mathrm{min})$. Nevertheless, interesting suggestions have been pushed forward that can be put to the test of microscopic simulations. 

\par 

Here we propose a parameterized formulation of the inverse-Bremsstrahlung electron-ion collision frequency with six constants, $(\bm{C_\mathrm{abs}}, \bm{\eta}, \bm{\epsilon_\ell}, \bm{C_\ell}, \bm{\eta_\ell}, \bm{\delta})$ which correspond to variations found in the literature, to be adjusted by CMDS
\begin{align}
\nu_{ei}^{IB}&=\bm{C_\mathrm{abs}}\,\nu_0[n_e, T_\mathrm{eff}(\bm{\eta}), Z]\,\ln(\Lambda_{ei}^{IB}),\label{eqnuei}\\
\nu_0[n_e, T_e, Z] &= \dfrac{4 \sqrt{2 \pi} \,e^4}{3 \, \sqrt{m_e}\, (4 \pi \epsilon_0)^2} \dfrac{n_e \, Z}{(k_B T_e)^{3/2}},\label{nu0}\\
T_\mathrm{eff}(x)&=T_e+x\,\,m_e\,v_E^2/\kb,\label{eq:teff}\\
\Lambda_{ei}^{IB}&=\left[\bm{\epsilon_\ell}+\bm{C_\ell}\,\frac{4\,\pi\,\varepsilon_0^{\frac{3}{2}}\,(\kb T_\mathrm{eff}(\bm{\eta_\ell}))^{3/2}}{Z\,e^3\,\sqrt{n_e}}\right]\,\left(\frac{\omega_p}{\omega}\right)^{\bm{\delta}},\label{eqnueifin}
\end{align} where 
\begin{align}
v_E^2=\left(\frac{e \tilde{E}}{m_e\omega}\right)^2=\frac{e^2\, (I\,\lambda^2)}{2\,\pi^2\,\varepsilon_0\,c^3\,m_e^2}\label{eq:vE}
\end{align} is the quiver velocity which is the maximum velocity of the oscillating motion of free electrons due to the electric field time variation assumed to be monochromatic and linearly polarized of the form $\bm{E(t)}=\tilde{E}\,cos(\omega\,t+\Phi)\,\bm{n}$ where $\bm{n}$ is a unit vector along the polarization direction (in \S\ref{sec:comp} other polarizations will be considered). 

\par 

The advantage of this parameterized formulation is that comparison with results of the literature can be made easier. In the following table (\ref{tab:model}), values of these six parameters for different references in the literature are compared in various regimes

\par 
\begin{table}[h!]
\begin{tabular}{l|c|c|c|c|c|c|c}
	Ref & Validity & $C_\mathrm{abs}$ & $\eta$ & $\epsilon_\ell$ & $C_\ell$  & $\eta_\ell$ &$\delta$ \\
	\hline\hline
	\cite{daw62,PF-Johnston-1973}& LiHf & 1 &0 & 0 & 1 & 0 & 1 \\
	\cite{silin65}& LiLf & 1 & 0  & 0 & 1 & 0 & 0 \\
	\cite{silin65}& LiHf (see \S\ref{sec:apA}) & 1 & 0  & 0 & 1 & 0 & 1 \\
	\cite{silin65}& HiLf & (a) & 0  & 0 & 1 & 0 & 0 \\
	\cite{jon82}& HiLf & (b) & 0 & 0 & 1 & 0 & 1 \\
	\cite{skup87}& Hf & 1 & 0 & 0 & 1 & 0 & 1 \\
	\cite{mulser20}& Hf & 1 & 0 & 0 & 1 & 1/4 & 1 \\
	\cite{brant03}& Hf & 1 & 1/6 & 0 & 1 & 0 & 1 
\end{tabular}
\caption{In this table, L and H stand for Low and High, i and f stand for intensity and frequency (for example, LiHf means Low intensity High frequency). This is an illustration of the fact that there are many different classical models in the literatures (different constants), and the list is not exhaustive. (a) grows like $(\ln\left(v_E/2 v_\mathrm{th}\right)+1)$ where $v_\mathrm{th}=\sqrt{\kb T_e/m_e}$, (b) grows like $(\ln\left(v_E/v_\mathrm{th}\right))$. More details on the constants reported in this table can be found in appendix \S\ref{sec:apB}.}\label{tab:model}
\end{table}

In the remainder of this article, the goal is to use molecular dynamic simulations to reach clear conclusions regarding values of these constants in order to discriminate between these models. 

\section{Numerical simulations dedicated to inverse bremsstrahlung absorption in the literature}\label{sec:num}

In the molecular dynamic simulations described in this paper, a plasma is described at the atomic level. Every particles of a plasma, electrons and ions, are described classically by their position and velocity and evolve as time goes by with Newton's first law.

\par 

The physical quantities we are interested in -- the electron-ion frequency, and in particular, the so called Coulomb logarithm that describes the manifestation of collective effects within the plasma -- depend upon two length scales that, in the context of weakly coupled plasma, are different by orders of magnitude. They are commonly called $b_\mathrm{min}$ and $b_\mathrm{max}$ and correspond, for the former, to the smaller distances of approach between electron and ions (a two-body effect) and, for the latter, to the Debye length (a collective effect). Therefore, in order to simulate these collisions and measure the Coulomb logarithm, it is of paramount importance to describe both scales precisely. Only molecular dynamics simulations allow for such a description. In the best case scenario, PIC simulations capture Debye length, $b_\mathrm{max}$, but under no circumstances can they capture $b_\mathrm{min}$. 

\par 

In the literature, numerical simulations dedicated to inverse bremsstrahlung fall into three categories : particle-in-cell simulations (PIC), Fokker-Planck simulations (FPS) and molecular dynamic simulations (either quantum molecular dynamic simulations, QMDS, or classical molecular dynamic simulations, CMDS). 

\par 

In the context of inverse Bremsstrahlung, FPS have mostly been used to assess the effect of the laser on the free electron distribution \cite{weng06, weng2009} which could turn from Maxwellian to super Gaussian of order 5 \cite{lang80} when the electron-electron collision frequency is much less that electron-ion collision frequency so that electron-electron collisions are not frequent enough to preserve the equilibrium shape. FP codes resolve the Boltzmann equation for single-particle velocity distribution function but it is well known that the collision source term in this equation depends upon the two-particle distribution function which in turn depends upon the three-particle distribution function and so on. This is the BBGKY hierarchy problem. In order to get a practical collision term it is mandatory, in this context of FP simulations, to make assumptions on the closure of this collision term \cite{lang80}. Therefore, these FP simulations are not able to let us gain full insight on microscopic quantities, such as collision-frequency or absorption, for their results rely heavily on the assumptions made on these very processes. PIC simulations are also plagued with the same issues. 

\par 

The first CMDS in the context of IB \cite{pfaz98} was carried out for strongly coupled plasmas and high intensity drive (non linear) because it is a situation that does not require to many particles (between 20000 and 40000  limited by the computational power back in 1998) to get proper results. Pfalzner and Gibbon were able to produce deformation of the free electron distribution as predicted by Langdon \cite{lang80}, though not in Langdon's condition which is $Z\gg 1$, and heating rate for $Z=1$ for a coupled plasma of $\Gamma=0.1$ and for $v_E/v_\mathrm{th}$ from 0.2 to 10. 

\par 

 In \cite{David2004}, David, Spence and Hooker carried out many CMDS (with approximately 16000 particles) resulting in several points of heating rate ($d T_e/d t$) versus laser intensity (from $10^{12}$ to $10^{17}$ W/cm$^2$) for different plasma states $(n_e, T_e)$ and compared their results to Polishchuk and  Meyer-Ter-Vehn's quantum model \cite{meyer94} but had to propose an alternative expression to get an agreement. We tested these numerical results against our parameterized model and showed that they are in agreement with our classical model.  
 
 \par
 
 Although not directly related to IB, CMDS reported in \cite{Dimonte2008, dim09} were the starting point of this work and are relevant to our parameterized model. These articles report on CMDS used to measure $\ln(\Lambda_{ei}^T)$ with repect to a parameter $g$ (which is nothing else than $\Gamma^{2/3}$ where $\Gamma$ is the plasma parameter). An almost perfect agreement was found when compared to the BPS theory \cite{bps05}. For the sake of completeness, we have found the same agreement with our own CMDS against Dimonte and Daligault's \cite{Dimonte2008, dim09} and against BPS (cf. Fig.\ref{fig:TRVR} in the present article). The analytical expression of the electron-ion collision frequency in the context of temperature relaxation is
 \begin{align}
 \nu_{ei}^T=\nu_0[n_e, T_e, Z]\,\ln(\Lambda_{ei}^T),\\
 \Lambda_{ei}^T=1+0.7\,\frac{4\,\pi\,\varepsilon_0^{\frac{3}{2}}\,(\kb T_e)^{3/2}}{Z\,e^3\,\sqrt{n_e}},
 \end{align} which, in the parameterized formalism, would correspond to $C_\mathrm{abs}=1$, $\eta=\eta_\ell=0$ (for there is no quiver velocity in this context), $\epsilon_\ell=1$ and $C_\ell=0.7$.
 
 \par
 
 In \cite{baal19}, Shaffer and Baalrud carried out velocity relaxation CMDS and found that the symmetry of charge was broken on the collision frequency at moderately to strong coupling. They found that for weakly coupled plasmas, CMDS with ions and positrons (positively charged electrons) and CMDS with ions and electrons (with statistically equivalent initial conditions) will evolve in a similar manner (statistically speaking). This is something we have found with or without electric field in our own CMDS. We have also checked that in the context of velocity relaxation, $\nu_{ei}^V(g)=\nu_{ei}^T(g)$ which means, in weakly coupled plasmas, from a CMDS stand point that $\Lambda_{ei}^T=\Lambda_{ei}^V$ (cf. Fig.\ref{fig:TRVR} in the present article).
 
\par 

In this article, we will only deal with Z=1 plasmas. Higher Z plasmas will be the subject of a future publication where the velocity distribution alteration, as it was first predicted by Langdon \cite{lang80}, will be central.

\section{Classical modelling of a two component plasma}\label{sec:mod}

\subsection{Equations of motion}

Describing a classical plasma consists in solving the classical equations of motions for all particles in the plasma submitted to their mutual Coulomb interactions. In the actual simulations, a soft core Coulomb potential as been used (instead of the $1/r$ potential) to avoid numerical problems, whose discussion is deferred to section \S\ref{sec:simset} (Simulations Settings), without affecting physical results. 

\par 

As long as the velocities involved are much less than the celerity of light $c$, the generated B-field is not strong enough to counteract onto the motion. In the context of inverse-Bremsstrahlung, one has to add the effect of an external varying electric field (corresponding to the laser). One can also neglect the electric field of the black-body radiation. Therefore, the equations of motions are as follow
\begin{widetext}
	\begin{align}
	m_e\,\frac{\mathrm{d}\bm{v}^{(\alpha)}(t)}{\mathrm{d}t}&=-\frac{e^2}{4\pi\varepsilon_0}\sum_{\beta\neq\alpha}\frac{\bm{n}_{\alpha\beta}}{|\bm{r}^{(\alpha)}-\bm{r}^{(\beta)}|^2}+\frac{Z\,e^2}{4\pi\varepsilon_0}\sum_{b}\frac{\bm{n}_{\alpha b}}{|\bm{r}^{(\alpha)}-\bm{R}^{(b)}|^2}-e\,\bm{E}(\bm{r}^{(\alpha)}(t), t),\label{eq1}\\
	m_i\,\frac{\mathrm{d}\bm{v}^{(a)}(t)}{\mathrm{d}t}&=\frac{Z\,e^2}{4\pi\varepsilon_0}\sum_{\beta}\frac{\bm{n}_{a \beta}}{|\bm{R}^{(a)}-\bm{r}^{(\beta)}|^2}-\frac{Z^2\,e^2}{4\pi\varepsilon_0}\sum_{b\neq a}\frac{\bm{n}_{a b}}{|\bm{R}^{(a)}-\bm{R}^{(b)}|^2}+Z\,e\,\bm{E}(\bm{R}^{(a)}(t), t),
	\end{align}   
\end{widetext} where $m_e$ and $m_i$ are the mass respectively of one electron and one ion (only one population considered), $Z$ is the charge of an individual ion ($+Z\,e$), $\bm{r}^{(\alpha)}$ is the position of an electron labelled $(\alpha)$ (with greek letters for electrons) and $\bm{R}^{(a)}$ is the position of an ion labelled by $(a)$ (with roman letters for ions). The unit vector $\bm{n}_{xy}$ is directed from particle $(x)$ to particle $(y)$ (where $x$ and $y$ can be the label of an electron or an ion). These are the exact equations taken into account in the forthcoming classical molecular dynamic simulations of a classical two components plasma. 

\subsection{Nondimensionalization of the equations of motion}\label{sec:nondim}

Owing to the mass difference between electrons and ions, $m_e\ll m_i$, the electrons can be assumed to collide on immobile ions to a very good approximation as long as \begin{align}
k\,T_i/m_i\ll k\,T_e/m_e.\label{hyp1}
\end{align} Therefore, in this limit, only equation (\ref{eq1}) needs to be dealt with. The characteristic geometrical length scale of a plasma is the typical distance between ions $\ell$, the characteristic velocity is the thermal velocity $v_\mathrm{th}$ and therefore the typical duration $\tau$
should be such that
\begin{align}
\ell&=n_i^{-1/3},\\
v_\mathrm{th}&=\left(\frac{\kb\,T_e}{m_e}\right)^{\frac{1}{2}},\\
\tau&=\frac{\ell}{v_\mathrm{th}},
\end{align} where $n_i$ is the ion density. Therefore, if one rescales lengths with $\ell$ and velocities with $v_\mathrm{th}$, every plasmas will end up with typical distances between ions equal 1 and velocity distribution of electrons width of 1 as well. What happens to other length scales, such as those related to \mbox{e-e}, \mbox{i-i} and \mbox{e-i} radial distributions functions, or time scales such as inverse of collision frequencies, depends entirely upon non dimensional parameters to be found in the remainder of this section. 

\par 

The electric field is assumed to be a pure monochromatic oscillation with pulsation $\omega=2\,\pi\,c/\lambda$ (where $\lambda$ is the actual wave length of the laser) and it is further assumed to be spatially uniform. This is compatible with our goal to evaluate, in a numerical experiment, values of electron-ion frequency in IB processes which are local physical quantities. Assuming the electric field is uniform amount to saying that the vacuum celerity of light is infinite. Indeed, if $\omega$ is fixed and $c=+\infty$ therefore $\tilde{k}=\omega/c=0$ that is to say, the wave length in this limit of $c$, $\tilde{\lambda}=+\infty$. Have we had less particles in our CMDS we could have claimed that the actual laser wave length, $\lambda$, is much larger than the simulation domain size $L$ but this is not quite correct for $10^6$ particles when one explores electronic densities as low as $10^{18}$ cm$^{-3}$ for $L$ becomes as large as $500$ nm. Since we are not interested in spatial mean inhomogeneities due to the spatial structure of the electric field, we are interested in a quantity that does not depend upon $c$. Therefore, we will consider that the actual $\bm{E}(\bm{x},t)$ only depends upon $t$  within the restricted simulation domain (the electric field is spatially uniform within that simulation domain). This is the same hypothesis made by all theoretical developments to model inverse bremsstrahlung absorption \cite{landau1936, daw62,silin65,PF-Johnston-1973,jon82,skup87, mulser01, brant03}. Finally, the polarization can be either linear, elliptical or circular such that 
\begin{align}
\bm{E}(t)=E_1\,\cos(\omega\,t)\,\bm{n_1}+E_2\,\sin(\omega\,t)\,\bm{n_2}\, ,\label{defE}
\end{align} with  $\bm{n_1}$ and $\bm{n_2}$ two orthonormal vectors perpendicular to the direction of propagation. 

\par 

Therefore, if $R=\ell\,\tilde{R}$, $r=\ell\,\tilde{r}$, $v=v_\mathrm{th}\,\tilde{v}$, $t=\tau\, \tilde{t}$ (where $\tau=\ell/v_\mathrm{th}$) and $E=(2\,I/c\,\varepsilon_0)^{1/2}\,\tilde{E}$, eq.(\ref{eq1}) can be recast as
\begin{widetext}
	\begin{align}
	\frac{\mathrm{d}\bm{\tilde{v}}^{(\alpha)}}{\mathrm{d}\tilde{t}}=g_\mathrm{coul}\,\left[\frac{1}{Z}\,\sum_{\beta\neq\alpha}\frac{\bm{n}_{\alpha\beta}}{|\bm{\tilde{r}}^{(\alpha)}-\bm{\tilde{r}}^{(\beta)}|^2}-\sum_{b}\frac{\bm{n}_{\alpha b}}{|\bm{\tilde{r}}^{(\alpha)}-\bm{\tilde{R}}^{(b)}|^2}\right]+g_\mathrm{osc}\,\bm{\tilde{E}}(g_\omega\,\tilde{t}),\label{eq1bis}
	\end{align}
\end{widetext} where 
\begin{align}
g_\mathrm{coul}&=\frac{Z\,e^2\,n_i^{1/3}}{4\,\pi\varepsilon_0\,k\,T_e},\\
g_\mathrm{osc}&=\frac{e}{n_i^{1/3}\,k\,T_e}\sqrt{\frac{2\,I}{c\,\varepsilon_0}},\\
g_\omega&=\omega\,\tau.
\end{align} This very simple analysis shows that plasmas with different temperatures, densities, laser sources may look different but as long as they have the same $Z$, $g_\mathrm{coul}$, $g_\mathrm{osc}$ and $g_\omega$, they are similar, that is to say, they will evolve identically when rescaled by the right length ($\ell$), velocity ($v_\mathrm{th}$) and time ($\tau$) factor. 

\par 

If eq.(\ref{eq1bis}) is integrated with respect to nondimensional time $\tilde{t}$, the velocity evolves as
\begin{widetext}
	\begin{align}
{\tilde{v}}^{(\alpha)}(\tilde{t})={\tilde{v}}^{(\alpha)}(0)+g_\mathrm{coul}\,\int_0^{\tilde{t}}\,\left[\frac{1}{Z}\,\sum_{\beta\neq\alpha}\frac{\bm{n}_{\alpha\beta}}{|\bm{\tilde{r}}^{(\alpha)}-\bm{\tilde{r}}^{(\beta)}|^2}-\sum_{b}\frac{\bm{n}_{\alpha b}}{|\bm{\tilde{r}}^{(\alpha)}-\bm{\tilde{R}}^{(b)}|^2}\right]\,\mathrm{d}\tilde{\tau}+\frac{g_\mathrm{osc}}{g_\mathrm{\omega}}\,\bm{\tilde{E}}^{(-1)}(g_\omega\,\tilde{t}).\label{eq:vit}
\end{align} 
\end{widetext} The first term in the right-hand-side of the equation is still piloted by $g_\mathrm{coul}$ but the second term, concerning the electric field, is now piloted by $g_\mathrm{osc}/g_\mathrm{\omega}$ which is exactly $v_E/v_\mathrm{th}$.

\par

Therefore, with no laser, $I=0$ or $g_\mathrm{osc}=0$, all nondimensionalized physical quantities (such as a coulomb logarithm) should only depends upon $Z$ and $g_\mathrm{coul}$ (which is nothing else than the plasma parameter). This is indeed the case as reported from molecular dynamic simulations of temperature relaxation \cite{Dimonte2008,dim09} (where $g$ in these references is proportional to $g_\mathrm{coul}^{3/2}$) and from theoretical studies \cite{bps05}.

\par 

With radiation, meaning $g_\mathrm{osc}\neq 0$, eq.(\ref{eq:vit}) tells us that every nondimensionalized physical quantities that depends upon velocity distribution (such as collision frequencies or absorption rate), should depend upon $Z$, $g_\mathrm{coul}$ and $g_\mathrm{osc}/g_\mathrm{\omega}=v_E/v_\mathrm{th}$ and not upon $Z$, $g_\mathrm{coul}$ and $g_\mathrm{osc}$ as eq.(\ref{eq1bis}) might have suggested. This will be highlighted in \S\ref{sec:comp}.

\section{Molecular dynamic simulations of a TCP with LAMMPS}\label{sec:lammps}

\subsection{LAMMPS code}

LAMMPS \cite{lammps} stands for Large-scale Atomic/Molecular Massively Parallel Simulator.  It is a PPPM (Particle-Particle-Particle-Mesh algorithm to account for periodic domain) code that has been developed at Sandia National Laboratory (New Mexico) with computational cost $O(N\log N)$, where $N$ is the number of particles simulated, as opposed to  a PP (Particle-Particle) code with higher computational cost $O(N^2)$.

\par 

The computational cost of the simulations limits the simulation domain to a small fraction of the size of a real plasma. This is mitigated by simulating a cube with periodic boundary conditions in all three directions, which is a valid approximation provided the Debye sphere of a given particle does not intersect with the Debye sphere of its replicas.

\par 

That has two consequences. The first and most obvious one is the fact that a particle going out of the domain through a boundary is immediately reintegrated to the domain through the opposite boundary with the same velocity. This is the reason why the number of particles in the domain will remain constant as time goes by along with total energy. This is the characteristic of a microcanonical simulation. The second consequence has to do with the interaction of particles. For instance, if a particle A interacts with a particle B in the domain, it also interacts with its own infinite replicates (all the As) and the infinite replicates of particle B in the periodic domains. This infinite sum can be efficiently carried out by the technic of Ewald summation. The way this is handled in PPPM molecular dynamic simulations \cite{grieb07} is by optimizing this Ewald summation using a fine regular mesh in the CMDS domain that will be used to calculate the long range part the electric potential with periodic replicates using fast-Fourier-transform and the short range part of the potential by simply adding up the contributions of neighboring particles. 
 
\subsection{Simulations settings}\label{sec:simset}

The numerical domain is defined by a periodic box of size $L\times L\times L$ with $N_\mathrm{ion}$ ions of charge $+Z\,e$ and $Z\,N_\mathrm{ion}$ electrons of charge $-e$. Here, we have tested both like-charges simulations (positive electrons and positive ions, as in \cite{Dimonte2008}) and opposed-charges simulations (negative electrons and positive ions) and found no difference for weakly coupled plasmas as reported in \cite{baal19}. 

\par 

In CMDS, close encounters between negative electrons and positive ions, which are very unlikely for weakly coupled plasmas but can happen on very few occasions, would produce a non conservative energy event that would ruin the outcome of the simulation. In order to avoid these events, the potential used in our CMDS simulations with LAMMPS is a soft-core (SC) potential
\begin{align}
V_\mathrm{SC}(r)=\frac{q_1\,q_2}{4\pi\varepsilon_0}\frac{1}{\sqrt{r^2+a^2}}
\end{align} which behaves as a pure Coulomb potential when $r\gg a$ and goes to a finite value $q_1\,q_2/(4\pi\varepsilon_0\,a)$ when $r\ll a$ (corresponding to a vanishingly small value of the electric field derived from that potential when $r\ll a$). Other soft core potentials exist in the literature, such as $\frac{q_1\,q_2}{4\pi\varepsilon_0}\frac{1}{r}\,\left(1-e^{-r/a}\right)$, but they are equivalent for the most part. If the value of $a$ is small enough but non zero, it allows most particles to feel a pure Coulomb potential (because they are at a distance of one another much greater than $a$ most of the time) but for those few particles that venture too close to an opposite-charge particle (within a distance $a$) the SC potential leaves them fly-by (no interaction at distances much smaller than $a$). It drastically differs from the pure Coulomb potential that would increase and would capture both particles into such a small orbit around one another (with radius $\ll a$) that their velocity would skyrocket and violate the time-step limitation of the simulation. The value \begin{align}
a=0.1\,\,\mathrm{\AA},
\end{align} set in all simulations that will be presented hereafter, was found to be small enough : it has been checked that results presented do not depend on such a small value of $a$ up to at least $0.1$ $\mathring{\mathrm{A}}$ (this was studied in \cite{pand17}).

\par 

Our simulations are parameterized by the number of ions $N_\mathrm{ion}$ allowed to evolve ($10^6$ in our simulations), by $Z$, the degree of ionization of ions (one variety of ion with exactly one degree of ionization in our simulations, at variance with real life plasmas where there are several varieties of ions, each with possibly several degrees of ionization) and by $n_e$, the electron density. These three parameters defined the size $L$ of the numerical domain by
\begin{align}
L=\left(\frac{Z\,N_\mathrm{ion}}{n_e}\right)^{1/3}.
\end{align} 

Initially, the positions of electrons and ions are randomly distributed throughout the simulation domain with a uniform law. Moreover, velocities of these particles are also randomly distributed with a Maxwellian distribution, at $T_e$ for electrons and $T_i$ for ions. Of course, these initializations are not physical since the actual radial distribution function $g(r)$, also known as the pair correlation function, is not constant even in a weakly coupled plasma. Therefore, before each simulation, the code is launched for a buffer period during which particles equilibrates and spatial and velocity distributions converge towards their physical state. The spatial distribution of electrons and ions becomes such that the Coulomb potential between ions and electrons, $V_\mathrm{ei}$, reaches a minimum. Therefore, during this buffer period, $V_\mathrm{ei}$ decreases and the average electron kinetic energy increases accordingly. A quantity defined in terms of quadrupole moments of the plasma distribution is introduced in section \S\ref{sec:quad} in order to monitor the evolution of the spatial distribution quantitatively. 
\par
In order to accurately capture the physics of the plasmas of interest, simulations are constrained by a number of assumptions. First, relativistic effects are not included, so the electron temperature should not be too high ($k_B\,T_e\ll m_e\,c^2$=511 keV) which is always satisfied in ICF plasmas.

\par 

 Furthermore, for the periodic boundary condition not to affect the physics of the plasma, every particle should be screened from its replicas, i.e. the size of the domain should be greater than twice the Debye length of the particles. As long as $T_i < \bar{Z} T_e$, which will always be the case in the configurations we consider, the Debye length of the ions is smaller than that of the electrons, so the most constraining condition is $2 \lambda_D < L$, that is to say, $2(\varepsilon_0\,k_B\,T_e/n_e \,e^2)^{1/2}<(Z\,N_\mathrm{ion}/n_e)^{1/3}$ which translates roughly to 
 \begin{align}
 N_\mathrm{ion} > \frac{1}{2\pi\,(Z+1)^{3/2}}\,\frac{1}{g}.
 \end{align} The smaller $g$ (coupling parameter), the larger the number of particles in the simulation ($(Z+1)\,N_\mathrm{ion}$), the more computationally costly the simulation. One can then evaluate the cost of one time iteration depending upon the numerical scheme. In a nutshell, it scales like $O(N_\mathrm{ion}\log N_\mathrm{ion})\sim g^{-1}\log g$ for a PPPM code such as LAMMPS.
 
\par

The time step of a simulation must be such that for all particles and at all time, the variation of the acceleration of a particle between two time step should not vary more than a fraction ($\ll 1$) of its value or energy might be lost in the process causing spurious effects. For that matter, the most stringent constraint is exerted by electrons, which are the lightest and most mobile particles (compared to the heaviest ions). On average, they travel at velocities of order $v_\mathrm{th}\sim (k_B\,T_e/m_e)^{1/2}$. The distance of closest approach to one another or to an ion is given by $b_\mathrm{min}\sim e^2/4\pi\varepsilon_0\,k_B\,T_e$. The time step should then be of order $dt\sim b_\mathrm{min}/v_\mathrm{th}$, that is to say
\begin{align}
dt\sim \frac{e^2\sqrt{m_e}}{4\pi\varepsilon_0\,(k_B\,T_e)^{3/2}}
\end{align}

\par

The purpose of all our simulations is to precisely quantify the time of the exponential relaxation (either for temperature of for drift velocity) that is to say $\tau_{ei}^{p,E}=1/\nu_{ei}^{p,E}$. Therefore, simulations must last for a suitable fraction of $1/\nu_{ei}^{p,E}$. This is why, the number of iterations should scale like $N_\mathrm{iter}^{p,E}\sim 1/(dt\,\nu_{ei}^{p,E})$, that is to say
\begin{align}
N_\mathrm{iter}^{p,E}\sim \frac{(4\pi\varepsilon_0\,k_B\,T_e)^3}{n_e Z e^6}\sim Z\,g^{-2}.
\end{align} The computational cost of a simulation is therefore given by the product of the number of iteration, $N_\mathrm{iter}$, times the computational cost of one iteration. For a PPPM code such as LAMMPS, it scales like $Z\,g^{-3}\log g$. Therefore, the weaker the plasma coupling, the more costly the simulation. 

\section{Simulations without time varying external electric field}\label{sec:dim}

Before carrying out MD simulations dedicated to inverse bremsstrahlung (IB), we have tested the simulation suite on well established configurations. We have been able to reproduce existing results with our MD simulations in temperature relaxation \cite{Dimonte2008, dim09} and velocity relaxation \cite{baal19}. 

\subsection{Relaxation towards a physical initial plasma state monitored by the quadrupole moment of charges distribution}\label{sec:quad}

One of the key problem in molecular dynamic simulation is to obtain a physical initial plasma configuration. Velocities of each particles within a simulated plasma at local thermal equilibrium at temperature $T$ can easily be randomly generated following the Maxwell distribution $\exp(-\frac{m\,v^2}{2\,k_B\,T})$. Positions, on the contrary, are more difficult to generate. Charged particles within a plasma are not randomly distributed following a uniform distribution (where every position would be equally likely) since like charges tend not to get too close to each other, at variance with opposed charges. In order to quantify the deviation to spatial uniform distribution one uses radial distribution functions, $g_{ei}(r)$, $g_{ii}(r)$, $g_{ee}(r)$, defined by the distribution of distances, respectively, between each pair of electron-ion ($ei$), ion-ion ($ii$) or electron-electron ($ee$).

\par 

In practice, particles positions are initially distributed randomly and uniformly in a CMDS. Simulations (without laser) are then run for a certain duration $\tau_{eq}$ at the end of which particles reach their physical spatial distribution with the right radial distribution functions. It is suggested in the literature \cite{baal19} that $\tau_{eq}$ should be approximately $60\,\omega_{pe}^{-1}$ by inspecting the evolution of radial distribution functions as time goes by. 

\par 

Here, we describe a scalar quantity whose time evolution allows to precisely grasp the relaxation of the uniform distribution towards the physical distribution. Any positions distribution of particles $\{\bm{x}^{(\alpha)}\}_{\alpha\in[1,N]}$ (where $N$ is the number of particles in the CMDS, $\alpha$ the label of one given particle with charge $q_\alpha$) can be characterized by its multipoles, the first two being the dipole (rank one tensor) and the quadrupole (rank two tensor). The averaged dipole per particle, $D_i$, is defined by 
\begin{align}
D_i=\frac{1}{N_e}\,\sum_\alpha q_\alpha\, x^{(\alpha)}_i,
\end{align} and the averaged quadrupole per particle, $Q_{ij}$, is defined by
\begin{align}
Q_{ij}=\frac{1}{N_e}\,\sum_\alpha q_\alpha\, \left(x^{(\alpha)}_i x^{(\alpha)}_j-\frac{\delta_{ij}}{3}\,|\bm{x}^{(\alpha)}|^2\right).
\end{align} Hidden in these multipoles is the information of the position distribution which is exactly what one needs to construct a scalar that could be monitored as time unfolds to observe the relaxation of positions distribution.

\par 

For a two components plasma, one can show that the dipole $\bm{D}$ evolves as the position of the centre of mass of the electron. Indeed, since 
\begin{align}
D_i=-e\,\sum_{\alpha_e}  x^{(\alpha_e)}_i+Z\,e\,\sum_{\alpha_\mathrm{ion}} x^{(\alpha_\mathrm{ion})}_i
\end{align} and since the total momentum of the plasma can be set to zero, for it is a conserved quantity, then the centre of mass position of the plasma, \begin{align}
m_e\,\sum_{\alpha_e}  x^{(\alpha_e)}_j+m_i\,\sum_{\alpha_\mathrm{ion}} x^{(\alpha_\mathrm{ion})}_i,
\end{align} is constant and can be set to zero without loss of generality. Therefore,
\begin{align}
D_i=-\frac{e}{N_e}\,(1-Z\frac{m_e}{m_i})\,\sum_{\alpha_e}  x^{(\alpha_e)}_i.
\end{align} If $N_e$ is the number of electrons in the domain, the centre of mass of the electron, $\sum_{\alpha_e}  x^{(\alpha_e)}_i$ is roughly in the centre of the domain within non coherent thermal position fluctuations of order $\ell/\sqrt{N_e}$ decreasing as $N_e$ is increased. Therefore, no interesting information can be extracted from the only scalar that can be made out of the dipole $\bm{D}$ which is $|\bm{D}|^2$.

\par 

\begin{figure}[!h]
	\includegraphics[width=\linewidth]{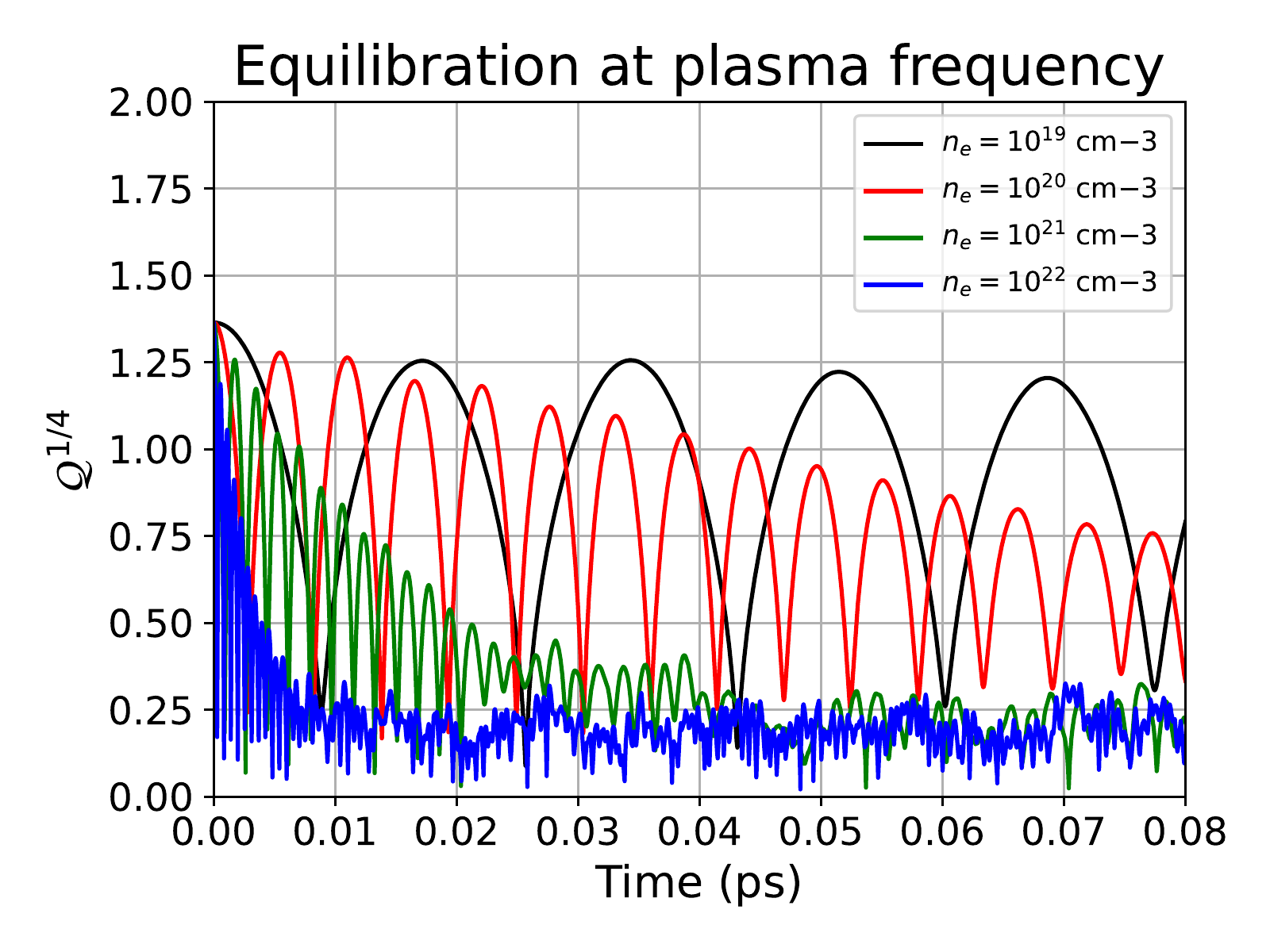}
	\caption{(color online) Evolution of the quadrupole, through ${\cal Q}^{1/4}$ (where ${\cal Q}$ is defined in eq.(\ref{eq:QQ})), with respect to time for four different plasma states, with $n_e=10^{19}$, $10^{20}$, $10^{21}$ or \mbox{$10^{22}$ cm$^{-3}$} initialized with random position distribution (uniform law) and Maxwellian velocity distribution corresponding to $T=$10 eV. The coherent oscillations observed for the four cases correspond to Langmuir waves at different plasma frequencies $\omega_p=\sqrt{n_e\,e^2/\varepsilon_0\,m_e}$ owing to different electronic densities. These coherent waves are progressively damped as the plasma evolves towards its statistically stationary state.}
	\label{fig:tq2}
\end{figure}

On the contrary, no such simplifications can be carried out on the quadrupole and the remaining question is how to construct an interesting scalar out of the quadrupole tensor ? The quadrupole is traceless by construction, therefore the trace of the quadrupole, which is the simple way to get a scalar (that is a rank 0 tensor) from a rank 2 tensor, is not an option. On the other hand, one can square the quadrupole and get the trace, $\mathrm{Tr}(\bm{Q}^2)=Q_{kj}Q_{kj}$, which is homogeneous to a length to the fourth power, that we can divide by the averaged length between electrons for instance ($n_e^{-1/3}$) at the fourth power to get a non-dimensional scalar, 
\begin{align}
{\cal Q}=n_e^{\frac{4}{3}}\,\mathrm{Tr}(\bm{Q}^2)=n_e^{\frac{4}{3}}\,\sum_{jk} Q_{kj}^2,\label{eq:QQ}
\end{align} that can be compared between different plasma state (cf. Fig.\ref{fig:tq2}).

The behavior depicted on Fig. \ref{fig:tq2} is typical of a plasma initialized ($t=0$) with a uniform spatial random distribution. For different initial electron densities, but for the same initial $T_e=10$ eV, the quantity ${\cal Q}$ undergoes a damped oscillation at the plasma frequency due to the fact that the initial spatial distribution is not physical and correspond to a perturbation with respect to its physical (equilibrium) counterpart. This perturbation is subsequently ($t>0$) damped out through plasma waves. Once these coherent oscillations reach a sufficiently low level, corresponding to the amplitude of the incoherent thermal fluctuations [that can be seen for $t>0.01$ ps on the blue curve ($n_e=10^{22}$ cm$^{-3}$) or for $t>0.04$ ps on the green curve ($n_e=10^{21}$ cm$^{-3}$) on Fig. \ref{fig:tq2}], one can be confident that the spatial distribution reaches an equilibrium state.

\par 

Once this physical state is reached, it constitutes the initial state of temperature relaxation (TR) simulations or velocity relaxation (VR) simulations, which were both carried out in order to compare our methodology with existing, well documented studies, and foremost, it constitutes the initial state of our inverse bremsstrahlung heating (IBH) simulations which are the core material of this publication.

\subsection{Temperature relaxation (TR) and velocity relaxation (VR)}\label{sec:VRTR}

The initialization previously described brings the electrons (at a density $n_e$) and the ions (at a density $n_i=Z\,n_e$) in a physical configuration (spatial distribution) at $T_e=T_i$ (velocity distribution). For (TR) simulations, one additionally requires that $T_i$ be different than $T_e$ in order to measure the relaxation rate to an equilibrium temperature. This can simply be done by only modifying the ions velocity distribution with $T_i > T_e$ or $T_i<T_e$. 

\par 

If the plasma, initialized this way, evolves freely, both temperatures will eventually equalize after a while when thermal equilibrium is reached. Both temperatures will reach the common equilibrium value following and exponential relaxation with time constant of order $1/\nu^{T}_{ei}$ defined by the energy electron-ion frequency
\begin{align}
\nu_{ei}^{T} = \frac{m_e}{m_i} \,\nu_0\,\ln\left(\Lambda_{ei}^T\right) \label{nueiE}
\end{align} characterizing the rate at which energy of an electron is significantly altered in its scattered motion through the plasma.

\par 

For a weakly-coupled plasma, the potential energy of the ee, ii and ei interactions are negligible compared to kinetic energy and, therefore, $n_e T_e + n_i T_i$ is constant throughout the evolution to a very good approximation. This can be used to eliminate $T_i$ from the evolution equation of the electron temperature 
\begin{align}
\dfrac{d T_e}{dt} =  - \nu_{ei}^{T} (T_e - T_i).
\end{align} From the fit of the solution of the resulting equation with the $T_e(t)$ relaxation time history in the CMDS, one can deduce, see \cite{Dimonte2008} for more details, the value of $\nu_{ei}^{T}$ and therefore that of $\ln\left(\Lambda_{ei}^T\right)$ for any given value of $g$ (compatible with the constraints of CMDS).

\begin{figure}[!h]
	\includegraphics[width=\linewidth]{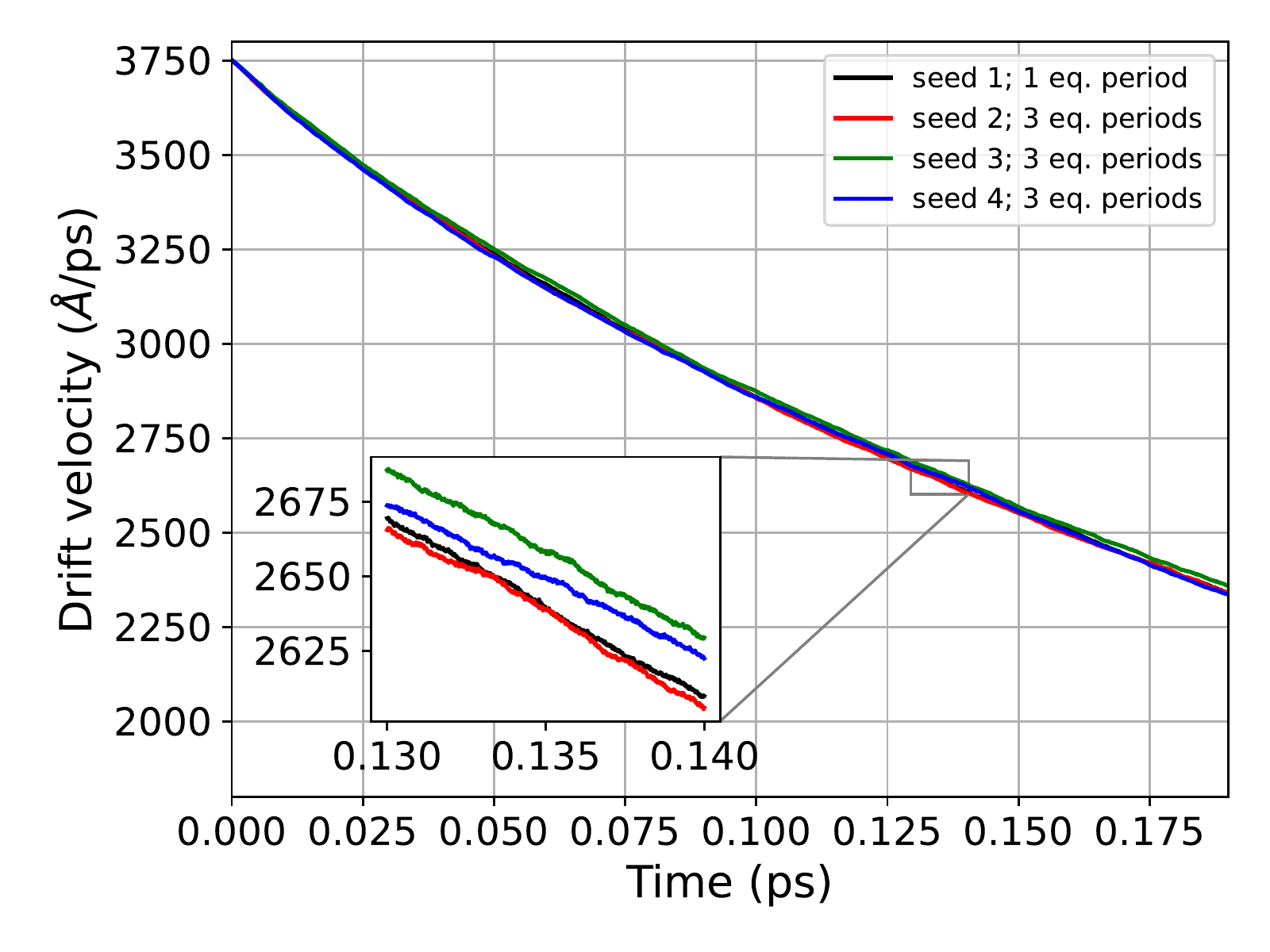}
	\caption{(color online) Evolutions of the drift velocity of a plasma initially at $n_e=10^{19}$ cm$^{-3}$, $T=$10 eV and $Z=1$, for 4 different equilibrations where the electron population has initially been given an ensemble velocity (drift velocity) where $v_d=0.1 v_{th}$. The drift velocity is decreasing in the same way for the 4 different initializations.}
	\label{fig:drift}
\end{figure}

For (VR) simulations, both the relaxed spatial distributions and the velocity distribution for ions and electrons are left unchanged. Therefore $T_e=T_i$, but a drift velocity component, $\bm{V}_d$, is added to every electrons so that the initial velocity of any single electron in a VR simulations is of the form $\bm{v}_\alpha=\bm{V}_d+\bm{v}^\prime_\alpha$ ($\alpha$ being the label of the electron under consideration) where $\bm{v}^\prime_\alpha$ is distributed according to a Maxwellian of temperature $T_e$. Therefore, in velocity space, the distribution of the electrons in the plasma is a Maxwellian shifted by $\bm{V}_d$ from the origin.

\par

With such an initialization, the electrons will flow through the web of ions with a drift velocity that will exponentially decrease as time goes by following
\begin{align}
\frac{\mathrm{d}\bm{V}_d}{\mathrm{d}t}=-\nu^V_{ei}\,\bm{V}_d,\label{eqVR}
\end{align} where 
\begin{align}
\nu_{ei}^{V} = \nu_0\,\ln\left(\Lambda_{ei}^V\right),\label{nueip}
\end{align} which characterizes the rate at which the momentum of an electron is significantly modified by its motion through scattering particles (electrons or ions) in the plasma (cf. Fig. \ref{fig:drift}).

\par

The comparison between (\ref{nueiE}) and (\ref{nueip}) shows a factor $m_e/m_i$ between $\nu_{ei}^T$ and $\nu_{ei}^V$. This is due to the fact that a single collision of an electron (small mass) against an ion (large mass, $m_i\gg m_e$) is enough to affect momentum (possible scattering of the electron in any direction) with almost no kinetic energy variation (elastic scattering) whereas it would require $\approx m_i/m_e > 2000$ collisions with ions to modify appreciably its kinetic energy. Molecular dynamics simulations of temperature relaxation are therefore more numerically costly for they need to be run for much longer time ($m_i/m_e$ times as long) than velocity relaxation. 

\par 

Extreme care should be brought to the choice of the initial $V_d$ as compared to $v_\mathrm{th}$. In order for the $\nu_{ei}^V$ measured in such simulation to be representative of the actual $\nu_{ei}^V$ of a plasma at local thermodynamic equilibrium (LTE) with $T_e=T_i=T$, the amplitude of the drift velocity $V_d$ should be much less than $v_\mathrm{th}$ for the velocity distribution of electrons to appear centred in the laboratory frame where ions have no ensemble averaged velocity. If $V_d$ were to be of order or greater than $v_\mathrm{th}$, not only would the velocity distribution of electrons in the same reference frame appreciably be shifted (by $V_d$) but it would gradually turn into a centred Maxwellian of larger width for the excess kinetic energy brought by the coherent motion ($V_d$) would dissipate, by collisional processes, into internal energy thereby increasing the electron temperature ($T_e$) significantly. 

\par 

The drawback is that taking $V_d\ll v_\mathrm{th}$ goes against signal to noise ratio. A molecular dynamic simulation being made of a limited number of particles, every averaged value, and the drift velocity is no exception, is subjected to statistical fluctuation, which is of order $v_\mathrm{fluc}\sim v_\mathrm{th}/\sqrt{N_\mathrm{part}}$ for drift velocities. Thus, in order to comply with the LTE constraint ($V_d\ll v_\mathrm{th}$) and still be able to get a sound measurement of the time history of $V_d$, which should not be buried under the statistical noise ($V_d\gg v_\mathrm{fluc}$), the drift velocity should verify $v_\mathrm{fluc}\ll V_d\ll v_\mathrm{th}$ to get a sufficient separation between $v_\mathrm{fluc}$ and $v_\mathrm{th}$. Two orders of magnitude, requires $N_\mathrm{part}>10^4$. All simulations presented in this article are carried out with $10^6$ ions and as many electrons. For all VR simulations, one used an initial 
\begin{align}
V_d=0.3\,v_\mathrm{th}
\end{align} which turned out to be a good compromise between LTE and statistical fluctuations. 

\par 

From the fit of the solution of that equation with the $\bm{V}_d(t)$ relaxation time history in the CMDS, one can deduce, see \cite{baal19} for more details, the value of $\nu_{ei}^{V}$ and therefore that of $\ln\left(\Lambda_{ei}^V\right)$ for any given value of $g$ (compatible with the constraints of CMDS). 

\par 

\begin{figure}[!h]
	\includegraphics[width=\linewidth]{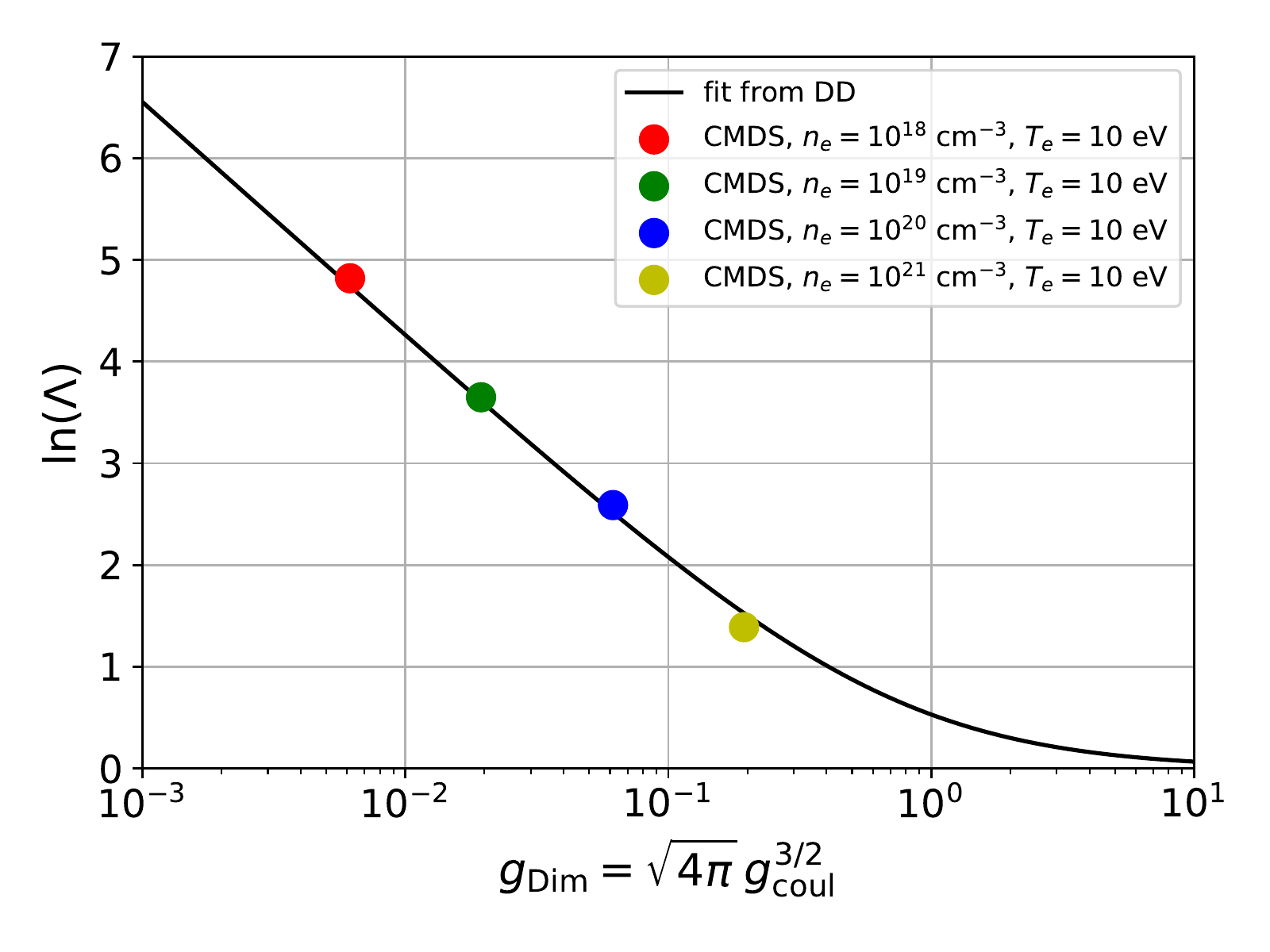}
	\caption{(color online) Plot of \mbox{$\ln\left(\Lambda_{ei}^E\right)=\ln(1+0.7/g)$} from Daligault and Dimonte \cite{Dimonte2008} in black solid line with colored points from our LAMMPS CMDS with $T_e=10$ eV and $n_e$ varied in such a way as to span the $g$ axis from $0.006$ to 0.2 for VR relaxation. The green point corresponds to the simulation described in Fig.\ref{fig:drift}. }
	\label{fig:TRVR}
\end{figure}

Clearly, our results on (Fig. \ref{fig:TRVR}) show very good agreement with results reported by \cite{Dimonte2008} also in agreement with theoretical results by \cite{bps05}. It is found that $\nu_\mathrm{ei}^{(V)}$ satisfies 
\begin{align}
	C_\mathrm{abs}&=1,\\ 
	\eta&=0, \\
	\epsilon_\ell&=1,\\
	C_\ell&=0.7, \\
	\eta_\ell&=0,
\end{align} which are the values reported by \cite{Dimonte2008} for $\nu_\mathrm{ei}^{(T)}$. This shows that, in weakly-coupled plasmas, molecular dynamics simulations agree with
\begin{align}
	\ln\left(\Lambda_{ei}^T\right)=\ln\left(\Lambda_{ei}^V\right).
\end{align} It is to be remembered, here (for $\nu_\mathrm{ei}^{(V)}$ and $\nu_\mathrm{ei}^{(T)}$), that $C_\mathrm{abs}=1$ whereas in our IB molecular dynamic simulations, even at low intensity, we shall measure that it is consistent with half that value, but we will come back to that in \S\ref{sec:comp}.

\section{Simulations with time varying external electric field dedicated to inverse bremsstrahlung heating}\label{sec:fin}

\subsection{Setup specific to IBH}

In our molecular dynamic simulations, the spatial variation of the electric field of the laser is not taken into account as mentioned in \S\ref{sec:nondim} and as it is the case in other such CMDS \cite{pfaz98, David2004}. This means that in no way can our CMDS provide any information about the dispersion relation for wave vector $k\neq 0$. 

\par 

In such simulations, the laser is mimicked by the uniform (in space) oscillating (in time) electric field. The oscillation frequency is such that $f_\ell=c/\lambda_\ell$. The effect of the electric field is to force a coherent ensemble motion of electrons (and ions) with an oscillating velocity whose maximum is the quiver velocity defined by (\ref{eq:vE}) (where $\omega=\omega_\ell=2\pi\,f_\ell$ is the laser pulsation and $\widetilde{E}$ the maximum amplitude of the electric field generated by the laser). 
\par

\begin{figure}[!h]
	\includegraphics[width=\linewidth]{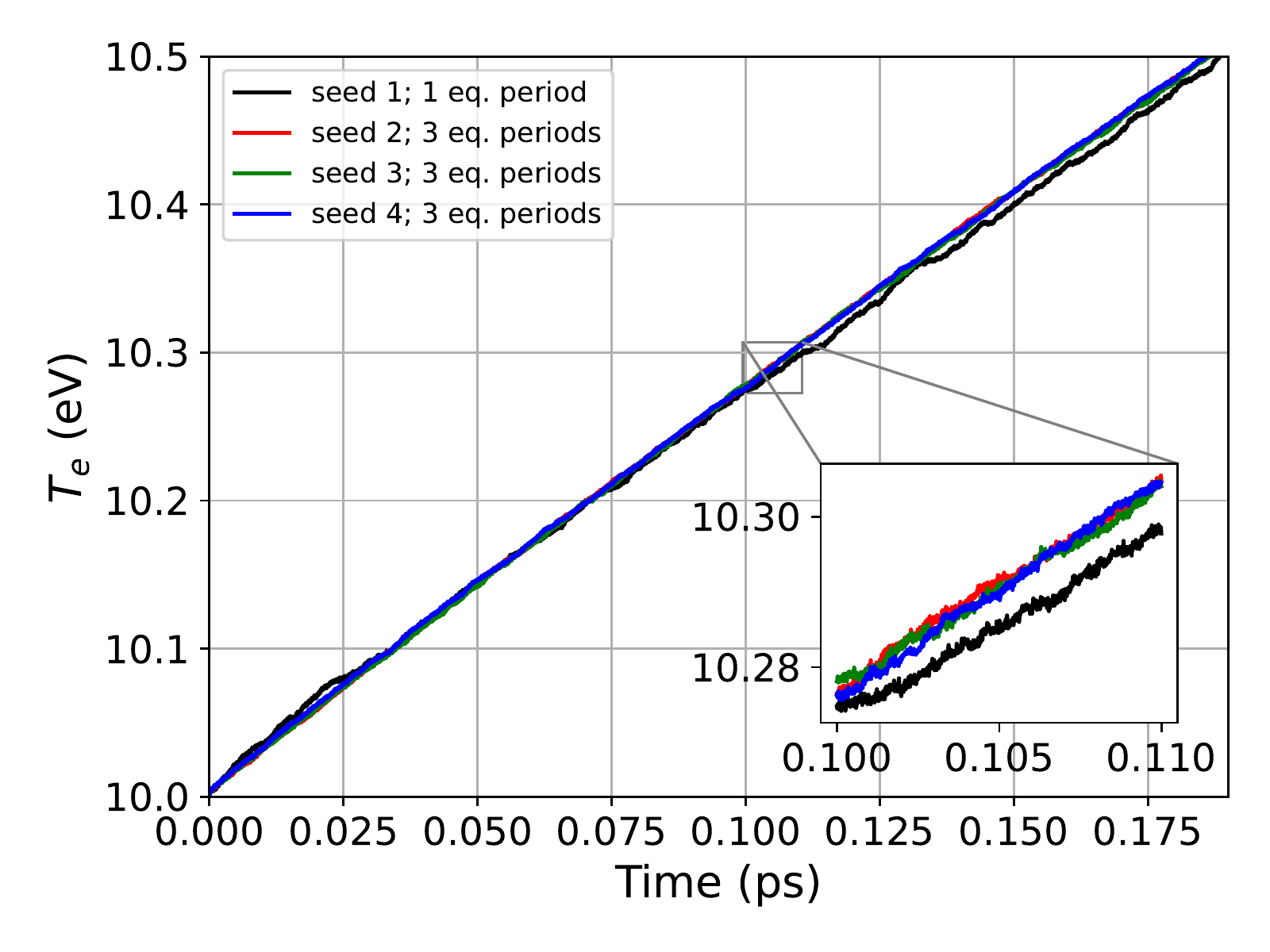}
	\caption{(color online) Evolution of the electronic temperature of a plasma initially at $n_e=10^{19}$ cm$^{-3}$, $T=$10 eV and $Z=1$, for 4 different equilibration processes, that is heated with a laser at $I=10^{14}$ W/cm$^2$ and $\lambda=351$ nm. The electronic temperature is increasing linearly, in the same way for the 4 different initializations, at a rate of 2.75 $\pm$ 0.05 eV/ps. Seed 1 to 4 correspond to different random series to generate the initial random velocities with maxwellian distribution. The number of equilibration periods correspond to the number of time the equilibration process was carried out. Although, "long time" ($\approx 0.025$ ps) fluctuation are clearly visible on the 1 equilibration period case, it does not affect the overall slope of the early linear increase of $T_e$ as time goes by compared to 3 equilibration periods.}
	\label{fig:terise}
\end{figure}

In the most general case, IBH simulations has to be initialized as VR simulations. Indeed, in the presence of an oscillating electric field, a coherent motion of the electrons (and ions) is set into play. There is an oscillating drift velocity created $\bm{V}_d(t)$ that verifies an equation similar to (\ref{eqVR}) where the electric field sets in as
\begin{align}
\frac{\mathrm{d}\bm{V}_d}{\mathrm{d}t}=-\nu^p_{ei}\,\bm{V}_d+\frac{e\,\bm{E}(t)}{m_e},\label{eqIBH}
\end{align} and with initial condition $V_d=0$. Assuming a varying electric field of the form (\ref{defE}), the general solution of (\ref{eqIBH}) at early time, when $V_d$ is still small (with respect to $v_\mathrm{th}$), is $\bm{V}_d(t)=\frac{e}{m_e\,\omega}(E_1\sin(\omega t)\,\bm{n}_1-E_2\cos(\omega t)\,\bm{n}_2)+\bm{V}^0_d$ where $\bm{V}^0_d$ is a possible constant vector of integration. Since initially $\bm{V}_d(0)=0$, it yields $\bm{V}^0_d=(e\,E_2/m_e\,\omega)\,\bm{n}_2$. When averaging the resulting $\bm{V}_d(t)$ over one laser period, it is found that $\langle\bm{V}_d(t)\rangle=\bm{V}^0_d=(e\,E_2/m_e\,\omega)\,\bm{n}_2$ since $\langle\cos(\omega t)\rangle=\langle\sin(\omega t)\rangle=0$. Therefore, once the electric field is set on, there is an immediate drift velocity in the most general case (when $E_2\neq 0$). 

\par 

This drift velocity relaxes with time but, as it does so, it although heats the electron population as described earlier in \S\ref{sec:VRTR} in the part concerning VR. This heating interferes with the one we want to monitor exclusively namely the inverse bremsstrahlung heating. In order to circumvent that issue, an initial drift velocity on the population of electron has to be enforced to compensate exactly for the one that will be triggered by the electric field. This is why, initially, one must set 
\begin{align}
\bm{V}_d(t=0)=-(e\,E_2/m_e\,\omega)\,\bm{n}_2.
\end{align}  

\par 

Once this is done, one can clearly appreciate the precision of the evolution of a typical electronic temperature history of one of our CMDS. In (Fig.\ref{fig:terise}), one displays the evolution of of a typical electron temperature time history, $T_e(t)$ (calculated by summing the individual kinetic energies of every electron in the reference frame of the oscillating electric field), for four different CMDS  with the same plasma and electric field parameters but originating from four different initializations such as described earlier in \S\ref{sec:quad}. Clearly, with $N_\mathrm{ion}=10^6$, the thermal fluctuation is barely noticeable which makes the measurement of the slope, $\mathrm{d}T_e/\mathrm{d}t$, very precise. This slope, also known as the heating rate, can be compared to the parameterized model eqs.(\ref{eqnuei})-(\ref{eqnueifin}) when included in eq.(\ref{eq:heatrate}).

\par 

\begin{figure*}[htb]
	\begin{center}
		\includegraphics[width=9cm]{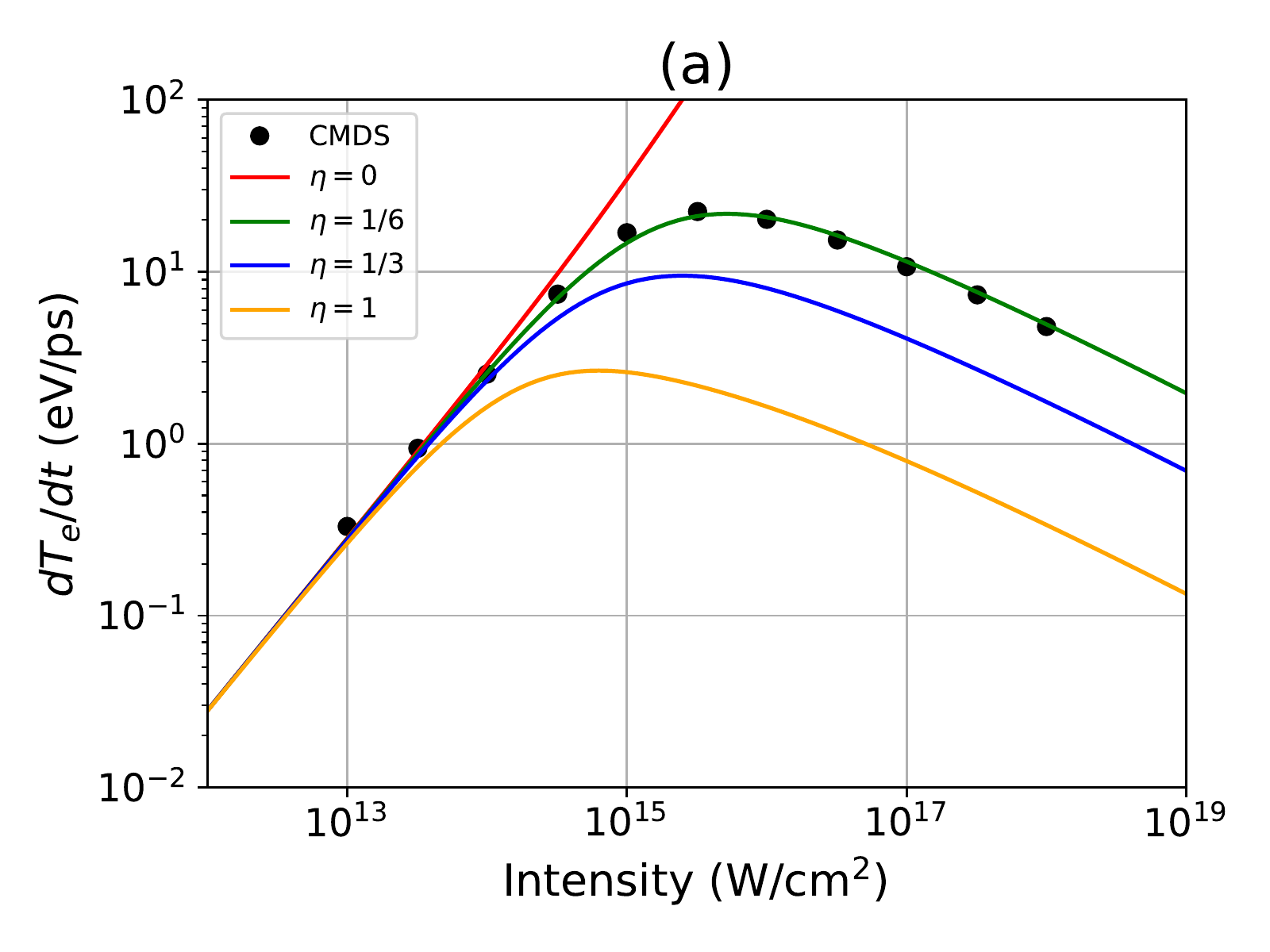}\includegraphics[width=9cm]{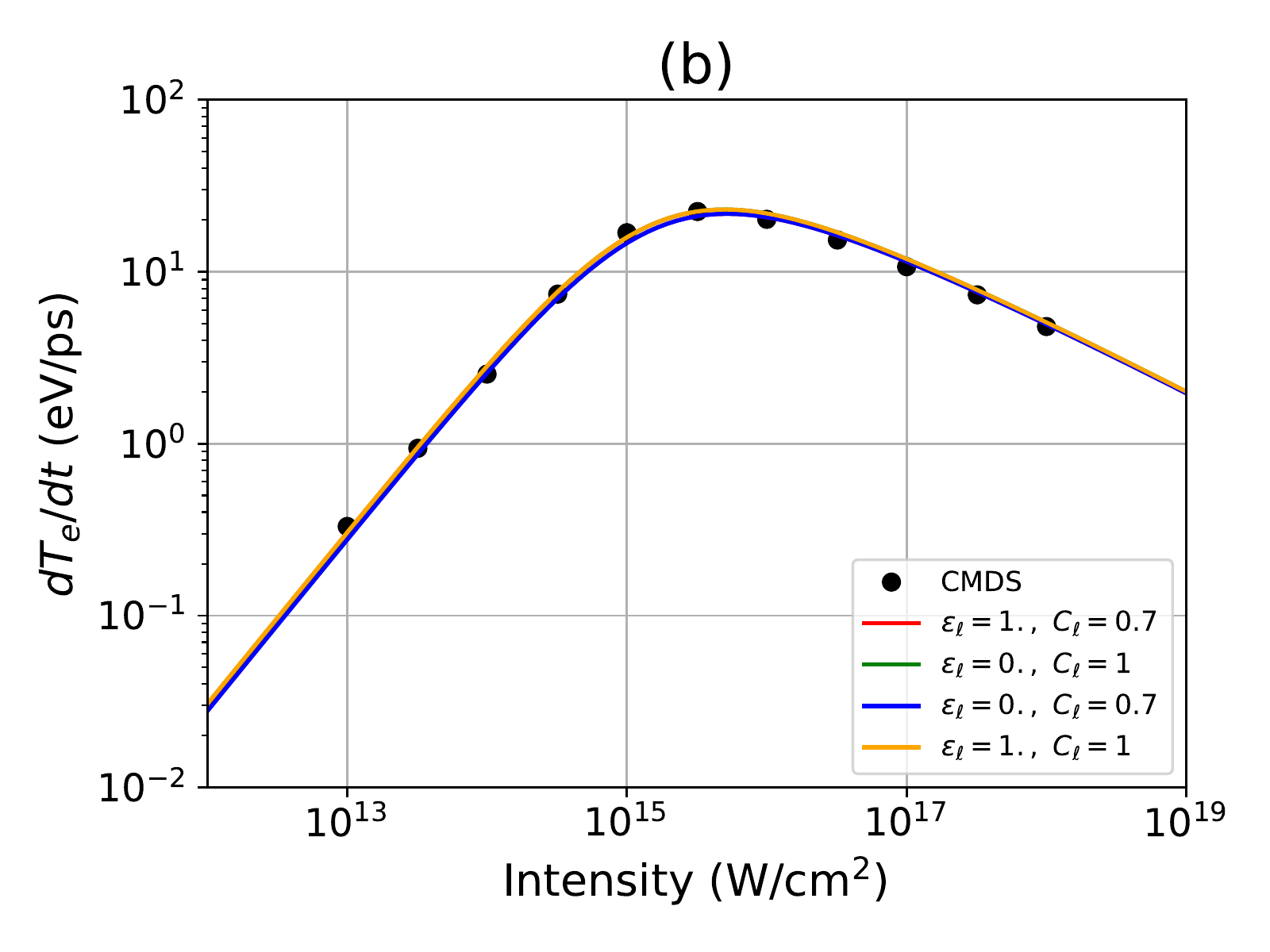}\\
		\includegraphics[width=9cm]{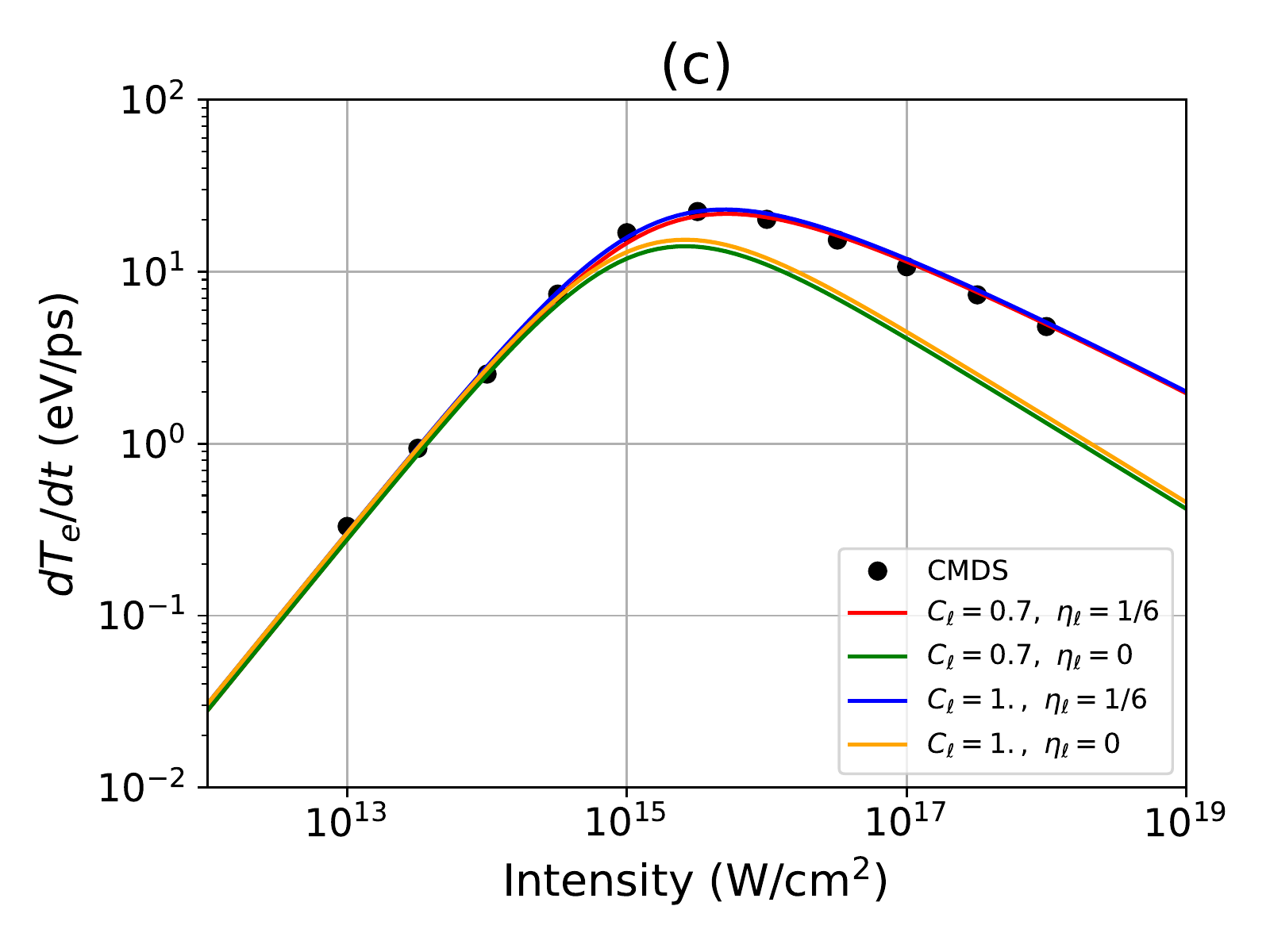}\includegraphics[width=9cm]{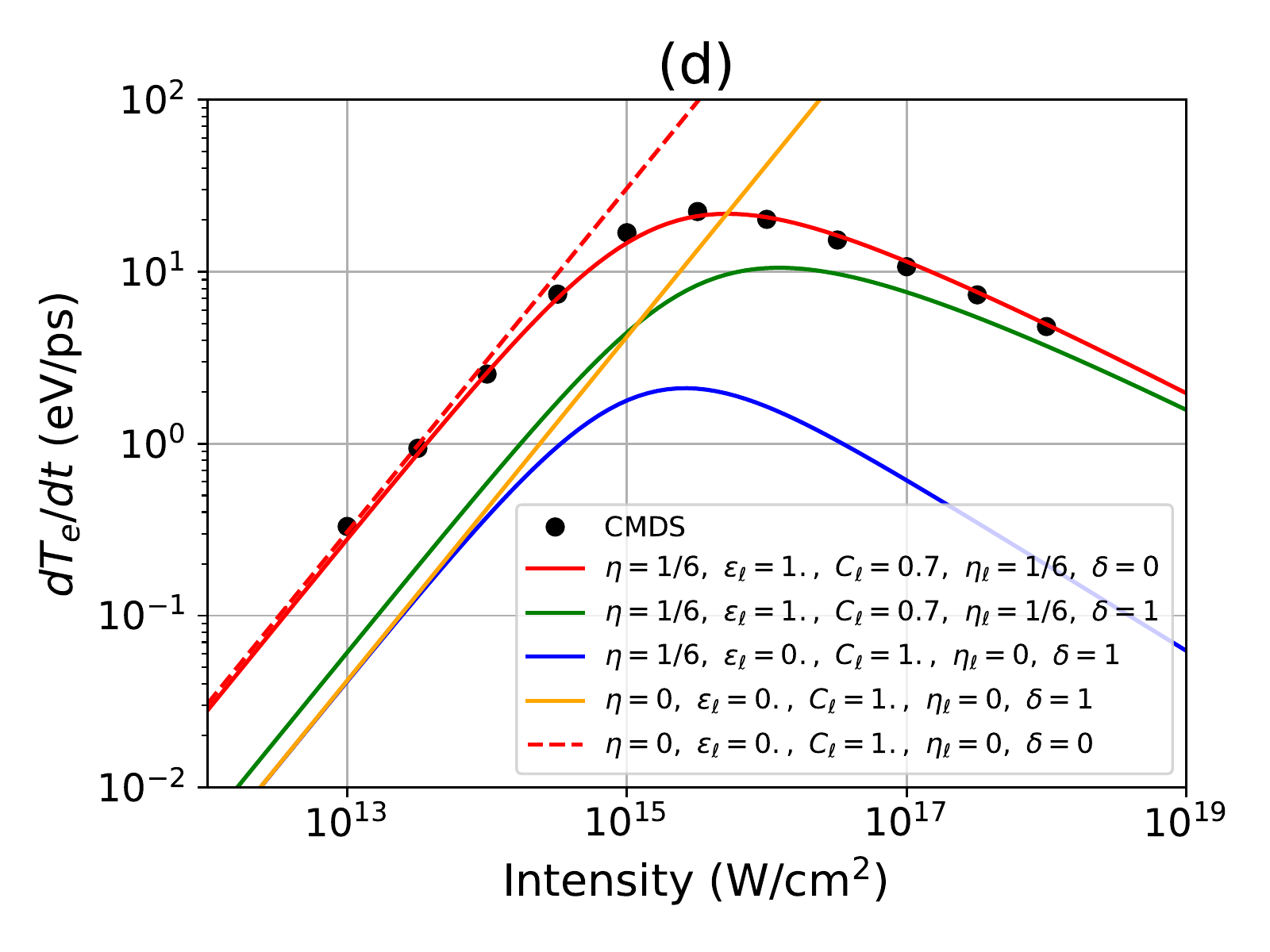}	
		\caption{(color online) Comparison of CMDS results (for $n_e=10^{19}$ cm$^{-3}$, $T_e=10$ eV and $\lambda_\ell=351$ nm) with the parameterized model for different adjustable constants variations. The important effect of $\eta$ is evaluated in figure (a). Figure (b) shows that the effect of $\varepsilon_\ell$ and $C_\ell$ are not very important within these reasonable limits. Figure (c) confirms the small effect of $C_\ell$ but shows the important effect of $\eta_\ell$. Finally, figure (d) shows the importance of $\delta$. The value $\delta=1$, put to the fore by most classical theories, is ruled out by CMDS.}
		\label{fig:ne19_mod}
	\end{center}
\end{figure*}

This is the important information we gather from these simulations in order to display CMDS measurements versus theoretical models such as those in \mbox{(Fig. \ref{fig:ne19_mod})} for instance. In this example, indeed, each black point corresponds to a full CMDS of $N_\mathrm{ion}=10^6$ ions and $N_\mathrm{elec}=10^6$ electrons over roughly $10^6$ iterations with time steps around $10^{-7}$ ps with $n_e=10^{19}$ cm$^{-3}$, $T_e=10$ eV with an oscillating electric field corresponding to $\lambda_\ell=351$ nm at intensities ranging from $10^{13}$ to $10^{18}$ W/cm$^2$. 

\par

The heating rate is calculated by a linear regression of the CMDS electronic temperature history between $t=0$ (corresponding to the instant the irradiation is turned on) and $t=t_\mathrm{fit}$, when the temperature deviates from a straight line by more than 0.1 $\%$ (calculated from a least squared fit). Therefore, what we call $\mathrm{d}T_e/\mathrm{d}t$ is the value at early time when (i) $T_e$ is still equal to its initial value (heating over a sufficiently long period may increase that temperature drastically) and (ii) the velocity distribution is still Maxwellian which may change as time unfold. This is a deliberate choice to be coherent with assumptions made by all theoretical investigations on IB heating \cite{landau1936, daw62,silin65,PF-Johnston-1973,jon82,skup87, mulser01, brant03} (Maxwellian distribution is one of these important assumptions).  

\begin{figure*}[htb]
	\includegraphics[width=9cm]{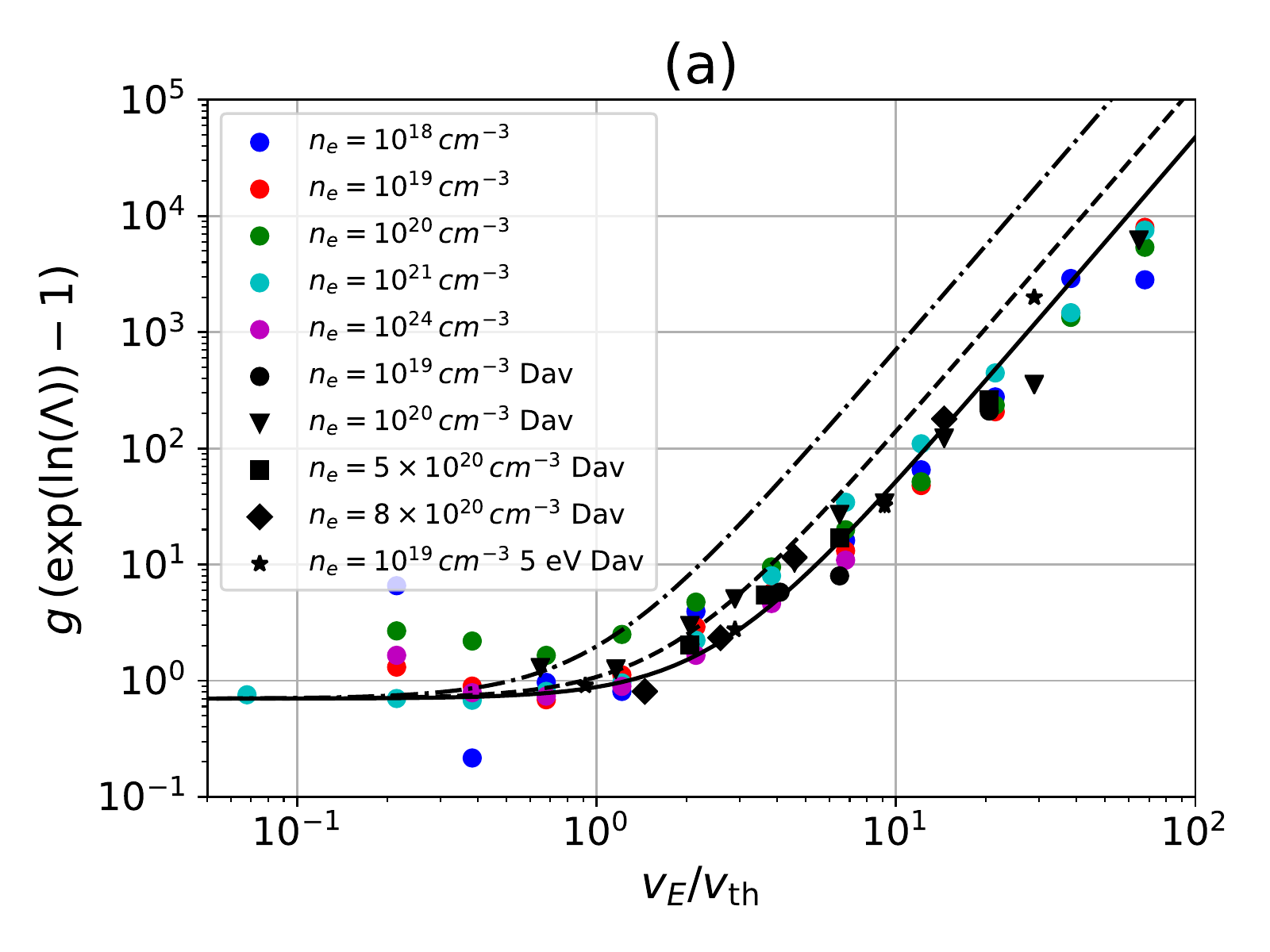}\includegraphics[width=9cm]{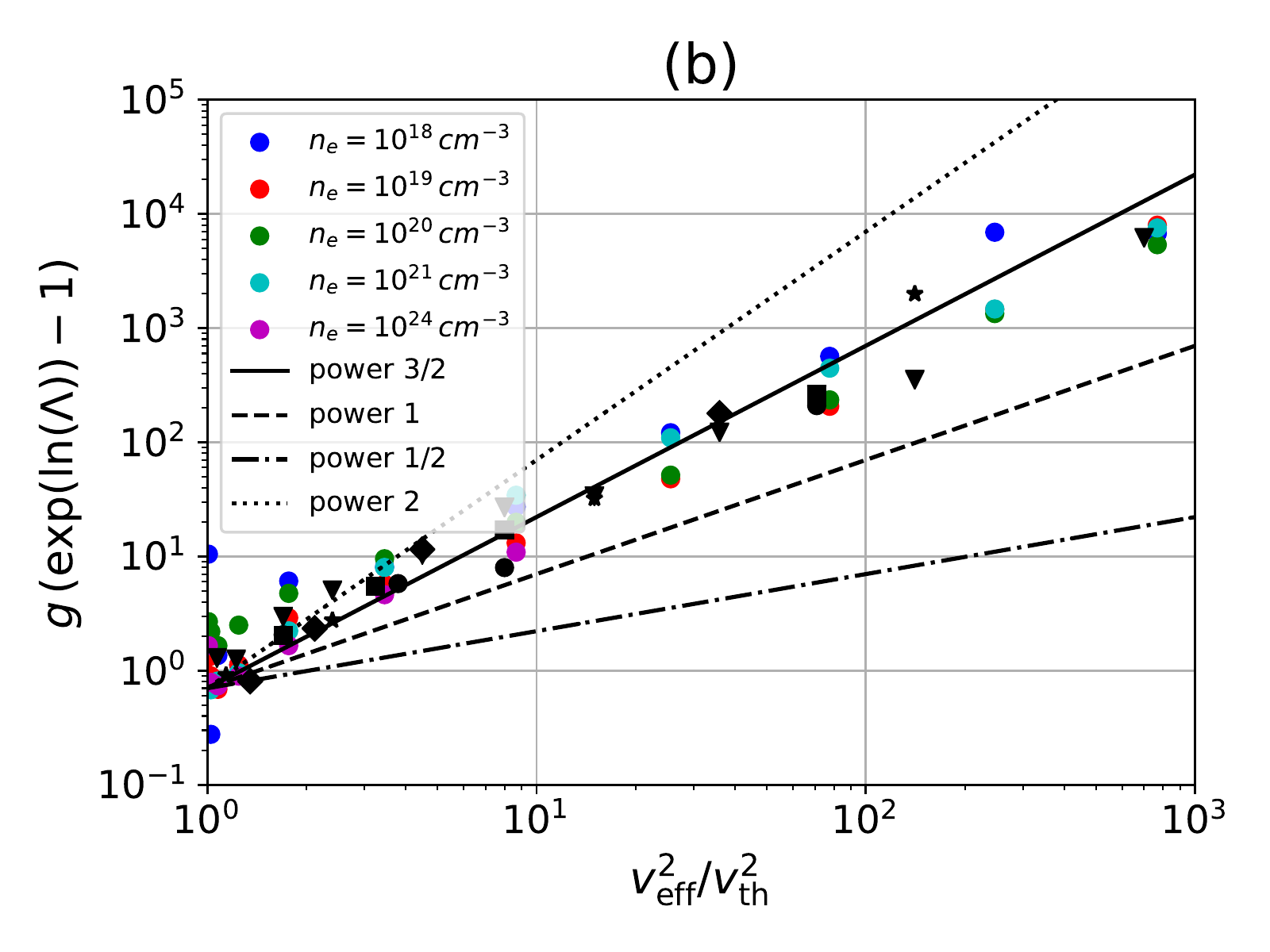}
	\caption{(color online) Points represent CMDS simulations from different origins. Colored points, with $\lambda=351$ nm, for CMDS from the present study and black points, with $\lambda=1060$ nm, from \cite{David2004}. On viewgraph (a), these points collapse with $g\,(\exp(\ln(\Lambda))-1)\propto (1+\eta_\ell\,v_\mathrm{eff}^2/v_\mathrm{th}^2))^{3/2}$ for $\eta_\ell=1/6$ in black solid line (dashed line corresponds to $\eta_\ell=1/3$ and dashed-point line corresponds to $\eta=1$). On viewgraph (b), molecular dynamics simulations points collapse with $g\,(\exp(\ln(\Lambda))-1)\propto (v_\mathrm{eff}^2/v_\mathrm{th}^2))^n$ with $n=3/2$, expected by Mulser \cite{mulser01, mulser20}, that cannot be mistaken with $n=0$ which is what most other past models advocate \cite{landau1936, daw62,silin65,PF-Johnston-1973,jon82,skup87}.}
	\label{fig:powerofI}
\end{figure*}

\subsection{Comparison between CMDS and the parameterized model}\label{sec:comp}

On one single set of CMDS, corresponding to a plasma state with $n_e=10^{19}$ cm$^{-3}$, $T_e=10$ eV, where the wave length of the electric field is fixed to $\lambda_\ell=351$ nm but where the intensity is varied, it is possible to adjust values of some constants of the parameterized model.

\par 

On each viewgraph in (Fig. \ref{fig:ne19_mod}), the 11 black points correspond to different values of the intensity evenly distributed in logarithmic scale (2 points per decade). Various settings of the parameterized model are plotted against our CMDS data. Viewgraph (a) demonstrates that the value $\eta= 1/6$ is, without a doubt, the only one compatible with molecular dynamics results. Viewgraph (c), also demonstrates that the value $\eta_\ell=1/6$ is the only one compatible with the CMDS data.  

\par 

This particular value $\eta=\eta_\ell=1/6$ is interesting for it is the precise value one would have found have we assumed that the coherent giggly motion of electrons due to the electric field oscillation was like an effective thermal motion that superimpose to the actual random thermal motion. This idea is not new for it is present in various earlier works such as \cite{brant03} and \cite{faehl78}. Since the actual velocity of a single electron is made out of a random thermal velocity $\bm{v}^\prime$ and of a coherent oscillation velocity $\bm{v}_E$, it can be recast as $\bm{v}=\bm{v}^\prime+\bm{v}_E$. From there, one can deduce $\bm{v}^2={\bm{v}^\prime}^2+\bm{v}_E^2+2\,\bm{v}^\prime\cdot\bm{v}_E$  and since $\bm{v}^\prime$ and $\bm{v}_E$ are uncorrelated, the average over one cycle of oscillation yields $\langle\bm{v}^2\rangle=3\,v_\mathrm{th}^2+v_E^2/2=3\,(v_\mathrm{th}^2+v_E^2/6)$ since $\langle{\bm{v}^\prime}^2\rangle=3\kb T_e/m_e=3\,v_\mathrm{th}^2$ and $\langle\bm{v}_E^2\rangle=v_E^2/2$ (average over 1 cycle of a squared sine is half its amplitude). One recognizes the factor $1/6$ between $v_\mathrm{th}^2$ and $v_E^2$.

\par 

Viewgraph (b) of (Fig. \ref{fig:ne19_mod}), shows that, within reasonable bounds ($\varepsilon_\ell=0$ and $C_\ell=1$ for most models \cite{landau1936, daw62,silin65,PF-Johnston-1973,jon82,skup87, mulser01, brant03} and $\varepsilon_\ell=1$ and $C_\ell=0.7$ for \cite{Dimonte2008, dim09}) it is difficult to discriminate between values of $\varepsilon_\ell$ and $C_\ell$ with our CMDS. Nevertheless, one has chosen to fix $\varepsilon_\ell=1$ and $C_\ell=0.7$ to be coherent with \cite{bps05, Dimonte2008, dim09} in the limit of vanishingly small intensities.

\par

Viewgraph (d) of (Fig. \ref{fig:ne19_mod}) displays several settings of the parameterized model with the best match to the CMDS data in red and with models from \cite{silin65,PF-Johnston-1973,jon82} in blue and \cite{skup87} in orange. The variation of the parameter $\delta$ clearly shows that $\delta =1$ is incompatible with CMDS data. The value $\delta=0$, not content with being in agreement with CMDS data, is also consistent from a theoretical point of view because one expect $\ln(\Lambda_{ei}^{IB})$ to converge towards $\ln(\Lambda_{ei}^{V})=\ln(\Lambda_{ei}^{T})$ in the vanishingly small intensity limit. In this this same limit, if one considers the case $\delta=1$, one is left with a dependency upon the laser pulsation in the coulombian logarithm whereas the oscillating electric field is almost turned off and that does not make sense.

\par 

It is possible to inspect the structure of $\ln(\Lambda_{ei}^\mathrm{IB})$ by dividing off the prefactor (\ref{nu0}) to the CMDS heating rate and taking the exponential in order to get $\Lambda_{ei}^\mathrm{IB}$. This is what was carried out and reported on (Fig. \ref{fig:powerofI}). On viewgraph (a), the parameterized model was plotted for three different values of $\eta_\ell$ (entering the expression of $\Lambda_{ei}^\mathrm{IB}$ in eq.(\ref{eqnueifin})) and clearly, even if the collapse of CMDS points is not perfect -- but it should be reminded that it is done on the exponential of $\ln(\Lambda_{ei}^\mathrm{IB})$, that is to say the exponential of CMDS results which increases drastically the uncertainty -- viewgraph (a) points toward $\eta_\ell=1/6$. Moreover, viewgraph (b) shows a clear behaviour of $\Lambda_{ei}^\mathrm{IB}$ in $(v_E/v_\mathrm{th})^3$ as explained by Mulser in \cite{mulser01, mulser20}. All points, including \cite{David2004}, collapse to the $I^{3/2}\propto v_E^3$ behaviour which strongly support that $T_e$ in the Coulomb logarithm should be replaced by $T_\mathrm{eff}$ of eq.(\ref{eq:teff}).

\begin{figure*}[htb]
	\begin{center}
		\includegraphics[width=9cm]{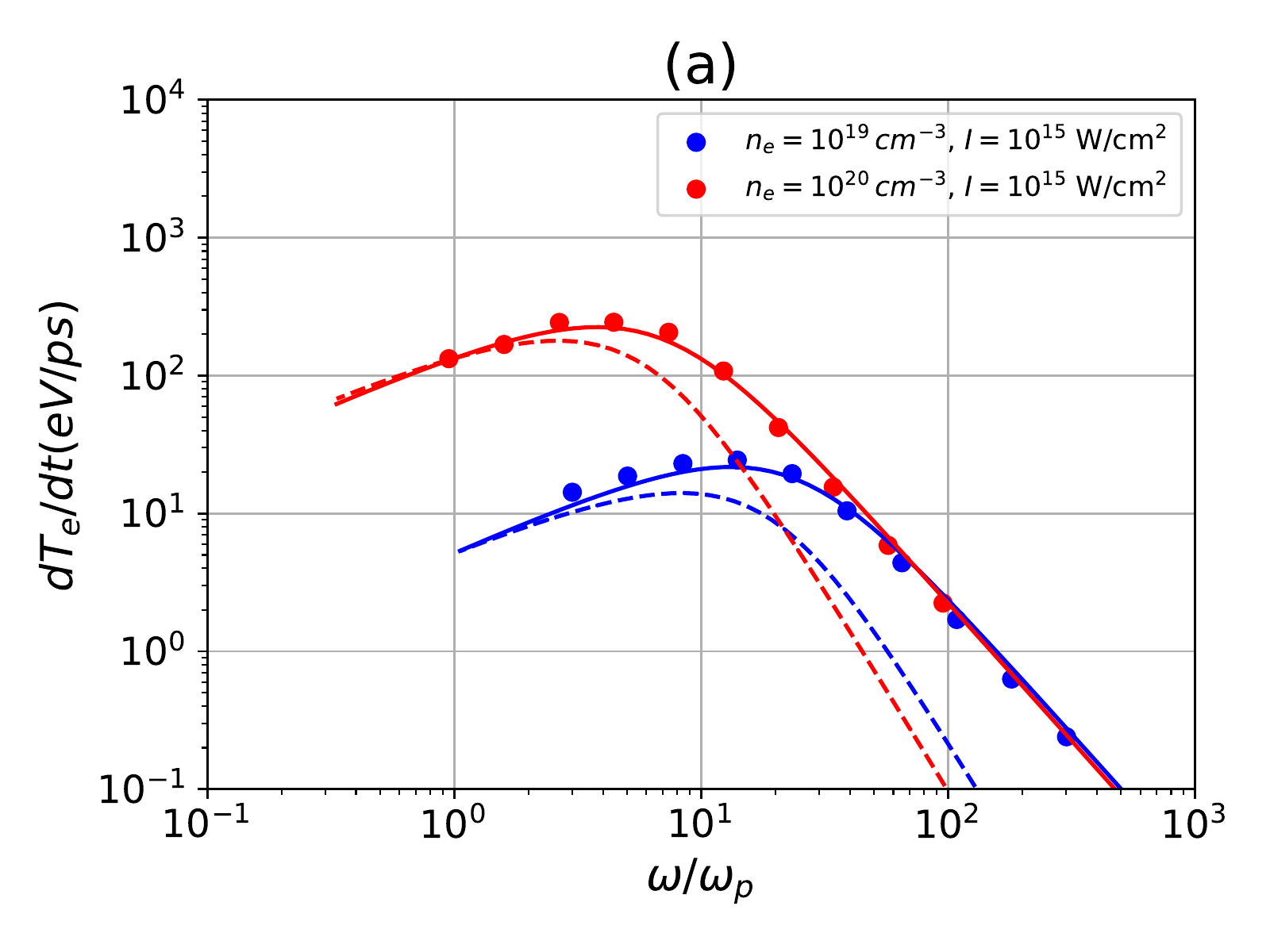}\includegraphics[width=9cm]{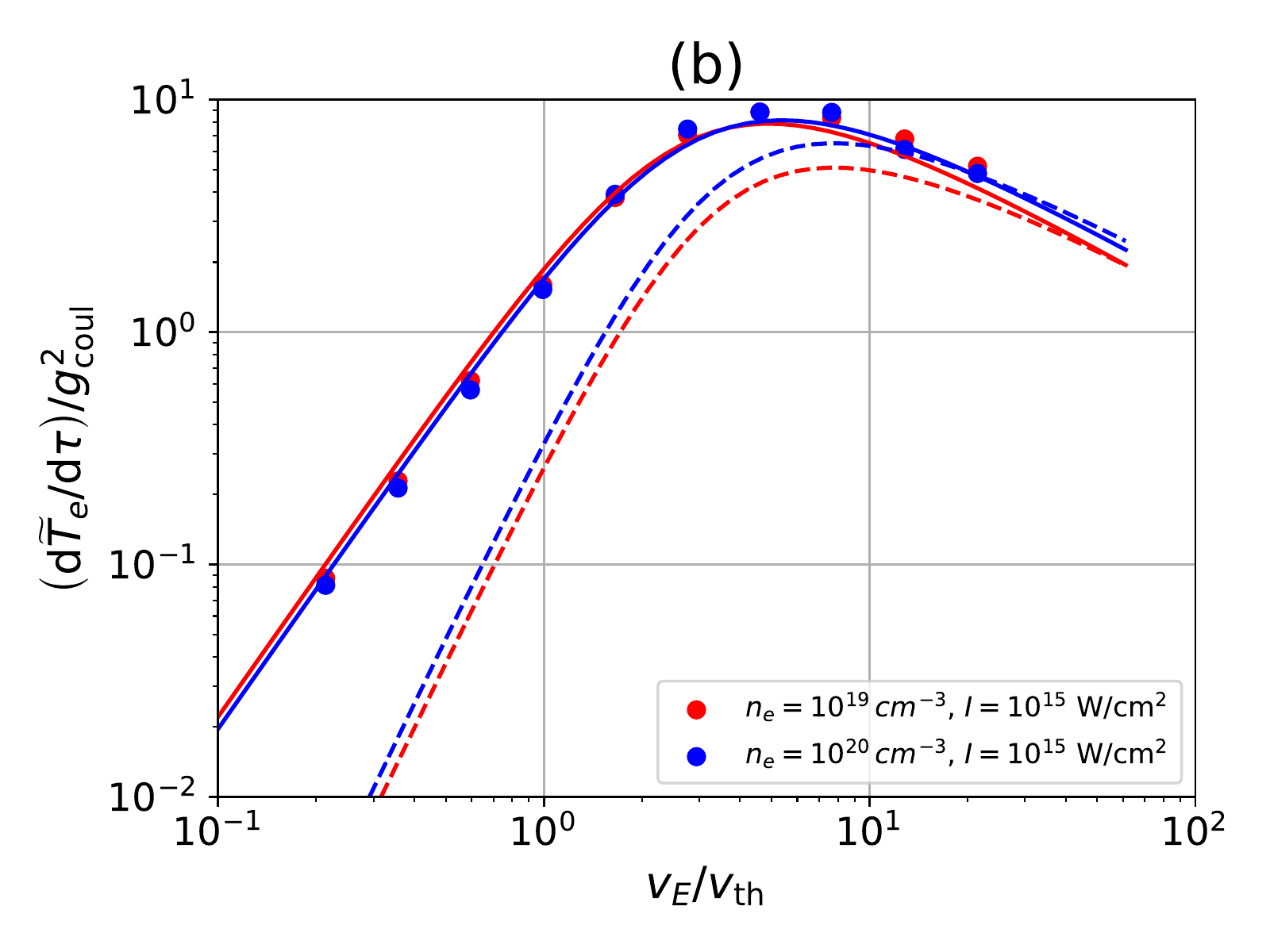}
		\caption{(color online) (a) Heating rate versus pulsation in unit of plasma pulsation ($\omega_p$) and (b) adimensionalized heating rate versus adimensionalized quiver velocity. Colored dots correspond to CMDS results at a fixed intensity of $10^{15}$ W/cm$^2$ for various laser pulsation (corresponding to wave lengths ranging from 35 to 3500 nm). Solid lines correspond to the best parameterized model with $\delta=0$ and dashed line with $\delta=1$.}
		\label{fig:neteint}
	\end{center}
\end{figure*}

\par 

It should be reminded here that only monochromatic oscillations of the electric field are considered. The question to be answered here is whether pulsation should enter the formal expression of the heating rate through the quiver velocity or could it steps in on its own, as reported by many publication starting from \cite{daw62, daw64}. This idea is encapsulated in the constant $\delta$ of the parameterized model that is non-zero in many publications as described in Table \ref{tab:model}. In order to get an answer to that interrogation, several CMDS have been carried out in the present study (Fig. \ref{fig:neteint}), maintaining the initial plasma state and the intensity of the irradiation constant and varying only its pulsation (for $\lambda$ ranging from 30 nm to 3000 nm). In viewgraph (a) results are presented as a function of $\omega/\omega_p$ and shows that all points considered here are under-critical. This set of simulations, again, strongly supports $\delta=0$ and viewgraph (b) shows that variations with respect to $\omega$ only shows up through $v_E/v_\mathrm{th}$ as expected from nondimensionalized equations in \S\ref{sec:nondim}. 

\begin{figure}[htb]
	\includegraphics[width=\linewidth]{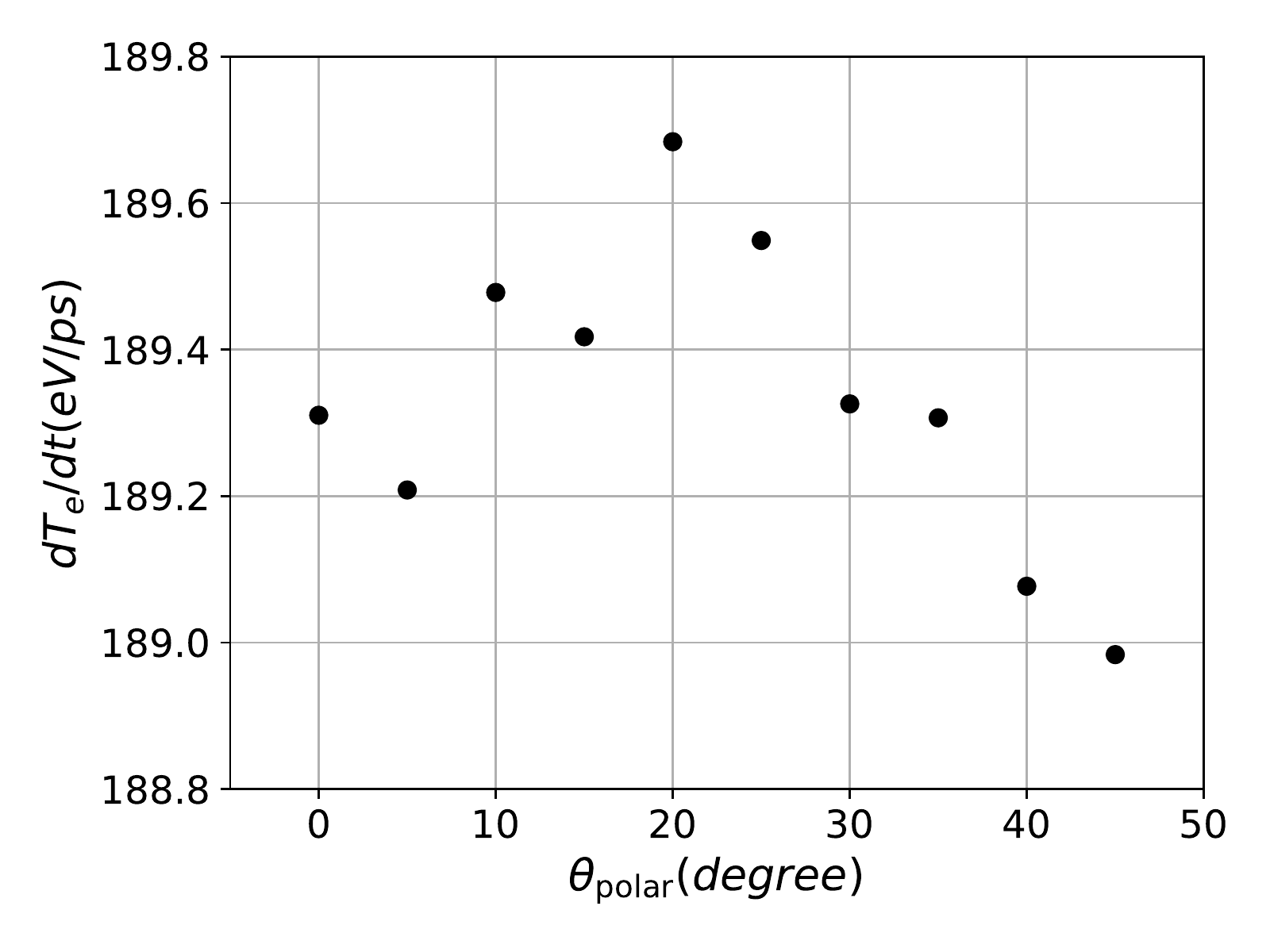}
	\caption{Effect of polarization on the heating rate for a plasma at $n_e=10^{20}$ cm$^{-3}$ and $T_e=10$ eV and for an intensity of \mbox{$10^{15}$ W/cm$^2$} at 351 nm. The variations observed are of statistical nature due to the finite number of particles in our simulations. As $\theta_\mathrm{polar}$ is varied, the heating rate does remain constant to within 0.15 $\%$ for CMDS carried out with $10^6$ ions.}
	\label{fig:polar}
\end{figure}

\begin{figure}[htb]
	\includegraphics[width=\linewidth]{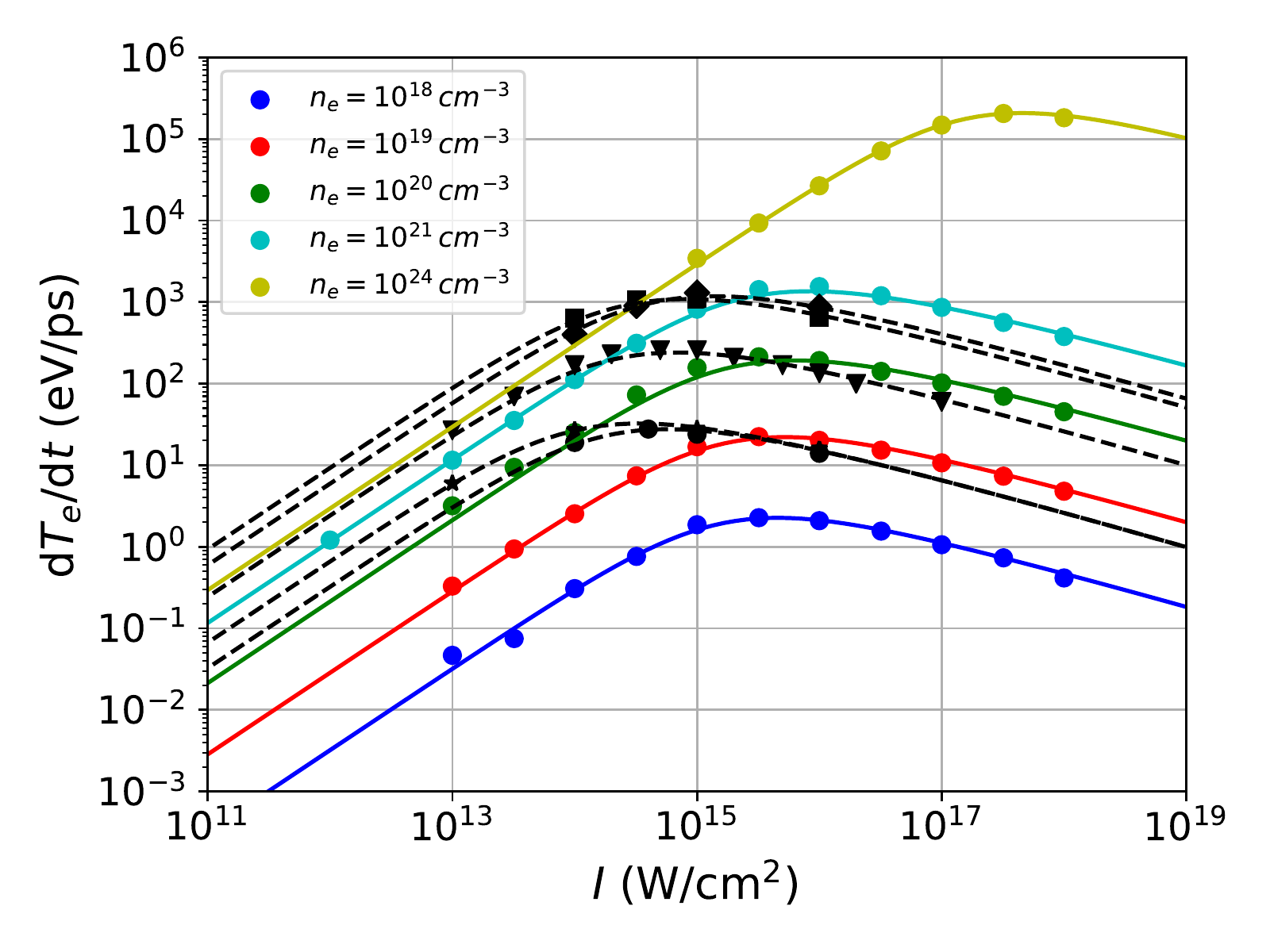}
	\caption{(color online) The colored dots represents initial $dT_e/dt$ versus laser intensity for plasmas initially at $T_e=10$ eV at different electronic densities ($10^{18}$, $10^{19}$, $10^{21}$, $10^{21}$ and $10^{24}$ $cm^{-3}$) extracted from our CMDS along with CMDS of David et. al. \cite{David2004} that are represented by black squares, triangles and stars. Solid colored curves represent the parametrized model for a wave length of 351 nm, with adjustable constants fixed to the values reported on eqs.(\ref{eqq1}-\ref{eqq2}), at the corresponding $n_e$ and $T_e$. Dashed black curves represent the parametrized model for a wave length of 1060 nm (which is used in \cite{David2004}), with adjustable constants fixed to the exact same values (eqs.(\ref{eqq1}-\ref{eqq2})).} 
	\label{fig:intens}
\end{figure} 


Effects of polarization on heating at a given intensity has also been investigated in the present study (Fig.\ref{fig:polar}). From eq.(\ref{defE}), the expression of $\langle E^2(t)\rangle$, where $\langle \,\rangle$ is the time average over one laser period, is $E_1^2 \langle cos^2(\omega t)\rangle+E_2^2 \langle sin^2(\omega t)\rangle$ and since $\langle cos^2(\omega t)\rangle=\langle sin^2(\omega t)\rangle=1/2$, it follows $\langle E^2(t)\rangle=(E_1^2+E_2^2)/2$. Let us define, $\theta_\mathrm{polar}$ such that $E_1=E_0\,\cos(\theta_\mathrm{polar})$ and $E_2=E_0\,\sin(\theta_\mathrm{polar})$. This yields $\langle E^2(t)\rangle=E_0^2/2$ whatever $\theta_\mathrm{polar}$. Therefore, by varying $\theta_\mathrm{polar}$ one spans all possible polarizations while keeping the intensity constant. A rectilinear polarization corresponds to \mbox{$\theta_\mathrm{polar}=0^\circ$}, a circular polarization corresponds to \mbox{$\theta_\mathrm{polar}=45^\circ$} and all other elliptical polarizations range between 0 and 45$^\circ$. The result in (Fig.\ref{fig:polar}) shows that there is no dependence of the heating rate upon polarization as seen from CMDS, at least, for low Z plasma. 

\par
\begin{table}[h!]
	\begin{tabular}{c|c|c|c|c|c}
		$n_e$ (cm$^{-3}$) & $T_e$ (eV) & $\lambda$ (nm) & I (W/cm$^{-2}$)& $C_\mathrm{abs}$ & data \\
		\hline\hline
		$10^{18}$ & 10  & 351 & $\mathrm{varI}$ & 0.48 & Fig.(\ref{fig:intens})\\	 
		$10^{19}$ & 10  & 351 & $\mathrm{varI}$ & 0.56 & Fig.(\ref{fig:intens})\\	 
		$10^{20}$ & 10  & 351 & $\mathrm{varI}$ & 0.60 & Fig.(\ref{fig:intens})\\	 
		$10^{21}$ & 10  & 351 & $\mathrm{varI}$ & 0.54 & Fig.(\ref{fig:intens})\\	 
		$10^{24}$ & 10  & 351 & $\mathrm{varI}$ & 0.44 & Fig.(\ref{fig:intens})\\		 
		$10^{19}$ & 10  & $\mathrm{var}\lambda$ & $10^{15}$ & 0.55 & Fig.(\ref{fig:neteint})\\	 
		$10^{20}$ & 10  & $\mathrm{var}\lambda$ & $10^{15}$ & 0.65 & Fig.(\ref{fig:neteint})\\
		\hline
	\end{tabular}
	\caption{In this table, values of $C_\mathrm{abs}$ are reported for different series of CMDS either by maintaining a constant laser wave length of 351 nm and varying the intensity from $10^{14}$ W/cm$^2$ to  $10^{18}$ W/cm$^2$ ($\mathrm{varI}$) or by maintaining the intensity constant at $10^{15}$ W/cm$^2$ and varying the wavelength from 35 to 3500 nm ($\mathrm{var}\lambda$).}\label{tab:stat}
\end{table}

The last point that needs to be addressed is the value of the overall factor $C_\mathrm{abs}$ in the parameterized model eqs.(\ref{eqnuei}-\ref{eqnueifin}). The way it is done in our numerical experiments is that this adjustable constant is calculated by fitting the parameterized model (with $\eta=\eta_\ell=1/6$, $\varepsilon_\ell=1$, $C_\ell=0.7$ and $\delta=0$) to each set of CMDS data at constant $n_e$, $T_e$. There were two kinds of data sets : those (i) maintaining the laser wave length constant and varying the laser intensity (Fig. \ref{fig:intens}) and those (ii) maintaining the laser intensity constant and varying the laser wave length (Fig. \ref{fig:neteint}). Values of $C_\mathrm{abs}$ were found to be scattered around 0.55 within a standard deviation of 0.07 as computed from table \ref{tab:stat}. Therefore, in a nutshell, that analysis as enabled to fix the six free parameters to the following values
\begin{align}
	C_\mathrm{abs}&=0.55\pm 0.07,\label{eqq1}\\
	\eta&=1/6,\\
	\epsilon_\ell&=1,\\
	C_\ell&=0.7,\\
	\eta_\ell&=1/6,\\
	\delta&=0.\label{eqq2}
\end{align}

\section{Conclusion}

The heating rate due to laser absorption by inverse bremsstrahlung is evaluated using the parameterized model described in eqs. (\ref{eqnuei}-\ref{eqnueifin}), which takes into account most models proposed in the literature (cf. Table. \ref{tab:model}) with appropriate parameter values $(\bm{C_\mathrm{abs}}, \bm{\eta}, \bm{\epsilon_\ell}, \bm{C_\ell}, \bm{\eta_\ell}, \bm{\delta})$. 

\par 

It was compared to results from classical molecular dynamic simulations (CMDS) carried out with LAMMPS \cite{lammps}. These simulations where shown to be in very good agreement with previous CMDS (with \cite{Dimonte2008, dim09} for velocity relaxation and with \cite{David2004} for inverse bremsstrahlung). CMDS presented here were carried out for a wide class of weak to moderate plasma coupling, for a wide variation of laser intensities (five orders of magnitude) for a fixed pulsation or for a wide variation of pulsations (two decades) for a fixed intensity. 

\par 

It was found that CMDS do not back up the rule, first put to the forth by Dawson and Oberman \cite{daw62, daw64} and derived by Silin \cite{silin65} for low intensity irradiation (as described in our \S\ref{sec:apA}), that $\omega_p$ in the expression of the Coulomb logarithm should be replaced by $\omega$ when $\omega\gg\omega_p$ or equivalently $\delta=1$ in our parametrized model. From a theoretical stand point, it tends to show that modeling the inverse bremsstrahlung heating should be done self-consistently by taking into account collective effects from first principles rather than assuming that they can be evaluated separately through a coulomb logarithm depending upon ad-hoc choices of short and long range characteristic lengths ($b_\mathrm{min}$ and $b_\mathrm{min}$).

\par 

Our classical molecular dynamic simulations of inverse bremsstrahlung heating are consistent with the parametrized model set to $\eta=\eta_\ell=1/6$, $\epsilon_\ell=1$, $C_\ell=0.7$, $\delta=0$ and $C_\mathrm{abs}=0.55$ which also matches previous CMDS by David et. al. \cite{David2004}. 

\appendix

\section{Silin's formula at low intensity}\label{sec:apA}

In his seminal paper \cite{silin65}, Silin did not derive explicitly a formula for the electron-ion frequency in the process of IB at low irradiation intensity. In \cite{Decker1994}, the authors said : "{\it He [Silin] presented a general expression
for the collision frequency in terms of complicated integrals. In the limit $v_E/v_\mathrm{th}\ll 1$ a closed form expression can be obtained and it is identical to that from the Dawson-Oberman model}". This is the calculation of this limit, from the {\it complicated integrals}, that we present in this appendix.

\par 

From eq.(3.8) of \cite{silin65}, instead of taking the limit of supra-critical plasma ($n_e\gg n_c$ or equivalently $\omega\ll\omega_p$ as in eq.(3.2) in \cite{silin65}) we take the low intensity limit, that is to say $\rho\ll 1$ in (3.9) which corresponds to $e\,E_0/m\omega\,v_T\ll 1$ in (3.8) of \cite{silin65}. 
\par

\begin{figure}[htb]
	\includegraphics[width=\linewidth]{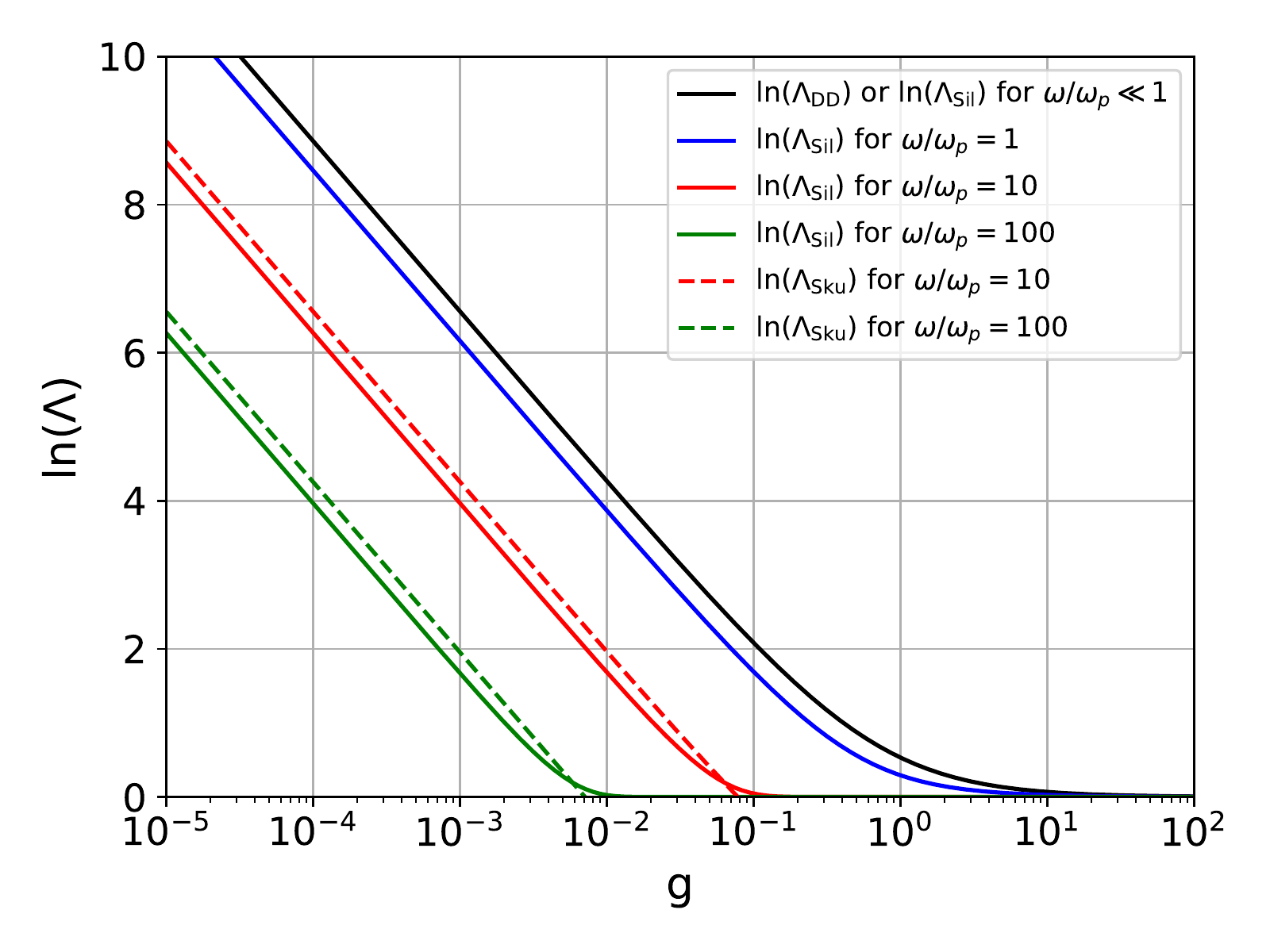}
	\caption{(color online) Plot of $\ln(\Lambda_\mathrm{Sil})$ using  \mbox{$\Lambda=\Lambda_\mathrm{DD}=(1+0.7/g)$} from Daligault and Dimonte \cite{Dimonte2008} for various values of $\omega/\omega_p$ and compared to Skupsky's prescription which consists in replacing $\omega_p$ by $\omega$ in $\Lambda=\frac{b_\mathrm{max}}{b_\mathrm{min}}=\frac{v_\mathrm{th}\,k_\mathrm{max}}{\omega_p}$ as soon as $\omega>\omega_p$. Clearly Skupsky's prescription is close to Silin results but none of these theoretical predictions match molecular dynamics simulation results.}
	\label{fig:logsilin}
\end{figure}

The electron-ion collision frequency in the process of inverse-Bremsstrahlung heating is given by 
\begin{widetext}
\begin{align}
\nu=\frac{16 Ni Z^2 e^2 \omega^3 m_e}{e\,E_0^3}\left[R\left(\frac{e E_0}{m\omega v_T},\frac{v_T k_\mathrm{max}}{\omega}\right)-R\left(\frac{e E_0}{m\omega v_T},\frac{v_T k_\mathrm{min}}{\omega}\right)\right],
\end{align}
\end{widetext} where the function $R$ is defined by
\begin{align}
R(\rho,x)&=\rho\int_0^x\mathrm{d}z\int_0^\infty\mathrm{d}y\,J_0(2\rho z \sin(y/2))\nonumber\\
&\times\left[e^{-x^2 y^2/2}+\frac{1}{2}\,Ei\left(-\frac{x^2 y^2}{2}\right)\right.\nonumber\\
&\left. -e^{-z^2 y^2/2}-\frac{1}{2}\,Ei\left(-\frac{z^2 y^2}{2}\right)\right].\label{eqR}\end{align} 
In the limit $\rho$ small, $J_0(x)$ can be expanded as $1-x^2$ at second order. If $J_0$ is replaced by its first order, that is 1, in (\ref{eqR}), then it is straightforward to show that $R=0$. Therefore, the first non vanishing term of the development of $R$ is due to the second order of the development of $J_0$. Therefore,  
\begin{align}
&R(\rho,x)\approx-4\rho^3\int_0^x\mathrm{d}z\,z^2\int_0^\infty\mathrm{d}y \sin^2(y/2))\nonumber\\
&\times\left[e^{-x^2 y^2/2}+\frac{1}{2}\,Ei\left(-\frac{x^2 y^2}{2}\right)-e^{-z^2 y^2/2}-\frac{1}{2}\,Ei\left(-\frac{z^2 y^2}{2}\right)\right].\label{eqR2}.
\end{align} Since
\begin{align}
&\int_0^\infty\mathrm{d}y \sin^2(y/2))\left[e^{-x^2 y^2/2}+\frac{1}{2}\,\mathrm{Ei}\left(-\frac{x^2 y^2}{2}\right)\right]\nonumber\\
&=\frac{\pi}{4} \mathrm{Erf}\left(\frac{1}{\sqrt{2}x}\right)-\frac{\sqrt{2\pi}}{4}\frac{e^{-\frac{1}{\sqrt{2}x}}}{x},
\end{align} the last integral, over $z$, in (\ref{eqR2}) can be recast as
\begin{align}
&R(\rho,x)\approx-4\rho^3\int_0^x\mathrm{d}z\,z^2\left(\frac{\pi}{4} \mathrm{Erf}\left(\frac{1}{\sqrt{2}x}\right)-\frac{\sqrt{2\pi}}{4}\frac{e^{-\frac{1}{\sqrt{2}x}}}{x}\right.\nonumber\\
&\left.-\frac{\pi}{4} \mathrm{Erf}\left(\frac{1}{\sqrt{2}z}\right)+\frac{\sqrt{2\pi}}{4}\frac{e^{-\frac{1}{\sqrt{2}z}}}{z}\right),
\end{align} and yields
\begin{align}
R(\rho,x)\approx-\frac{\rho^3}{12}\sqrt{\frac{\pi}{2}}\,\mathrm{Ei}\left(-\frac{1}{2 x^2}\right).
\end{align} One can then write the collision in the inverse Bremsstrahlung context as
\begin{align}
\nu=\frac{1}{2}\nu_0\,\left[\mathrm{Ei}\left(-\frac{\omega^2}{2 v_T^2 k_\mathrm{min}^2}\right)-\mathrm{Ei}\left(-\frac{\omega^2}{2 v_T^2 k_\mathrm{max}^2}\right)\right]
\end{align} and since $k_\mathrm{min}=2\pi/b_\mathrm{max}$ and $k_\mathrm{max}=2\pi/b_\mathrm{min}$ (where $b_\mathrm{max}$ and $b_\mathrm{min}$ are not deduced from first principle as in BPS but just cut-off of the theory), one can specified $2 v_T^2k_\mathrm{min}^2$ to be $\omega_p^2$ and then $2 v_T^2 k_\mathrm{max}^2=2 v_T^2 k_\mathrm{min}^2 (k_\mathrm{max}^2/k_\mathrm{min}^2)=\omega_p^2\,\Lambda^2$. The final result is 
\begin{align}
\nu=\nu_0\,\ln(\Lambda_\mathrm{Sil})
\end{align} where
\begin{align}
\ln(\Lambda_\mathrm{Sil})=\frac{1}{2}\,\left[\mathrm{Ei}\left(-\frac{\omega^2}{\omega_p^2}\right)-\mathrm{Ei}\left(-\frac{\omega^2}{\omega_p^2\,\Lambda^2}\right)\right]
\end{align} is displayed in (Fig. \ref{fig:logsilin}). The coulomb logarithm from Silin $\ln(\Lambda_\mathrm{Sil})=\ln(\Lambda)$ when $\omega\ll\omega_p$ (in the over-critical regime, which corresponds to eq.(3.12) in \cite{silin65}) and it goes to $\ln(\Lambda_\mathrm{Sil})=\ln(\Lambda\frac{\omega_p}{\omega}\,e^{-\gamma/2})$ when $\omega\gg\omega_p$ (in the super-critical regime, of interest to ICF) where $\gamma\approx 0.577$ is the Euler constant. It is interesting to note that if one write 
\begin{align}
\Lambda=b_\mathrm{max}/b_\mathrm{min}=v_\mathrm{th}\,k_\mathrm{max}/\omega_p,\label{lamb}
\end{align} the $\Lambda_\mathrm{Sil}=\Lambda\frac{\omega_p}{\omega}$ when $\omega\gg\omega_p$ amounts to replacing $\omega_p$ in (\ref{lamb}) by $\omega$ which is exactly what is done in Dawson-Oberman \cite{daw62} in eq. (26), in Johnston-Dawson \cite{PF-Johnston-1973} just below eq.(1b), in Jones-Lee \cite{jon82} above eq.(28) "the small wave number cut-off is $\omega/v_\mathrm{th}$", in Skupsky \cite{skup87} with its explicit prescription in eq.(3a) and in Mulser et. al. \cite{mulser01} in eqs.(15) and (16).  

\section{Details on the constants reported in table \ref{tab:model}}\label{sec:apB}

In Dawson-Oberman \cite{daw62} or Johnston-Dawson \cite{PF-Johnston-1973}, the collision frequency is provided in the low intensity ($v_E\ll v_\mathrm{th}$) and high frequency ($\omega\gg\omega_p$) limit, $C=1$ and $\ln(\Lambda_{ei}^{(IB)})=\ln(\Lambda\,\omega/\omega_p)-\gamma$ where $\gamma\approx0.577$ is the Euler constant. Skupsky \cite{skup87}, in its development about classic plasmas (as opposed to quantum), used similar Coulomb logarithm of the form $\ln(\Lambda_{ei}^{(IB)})=\ln(\Lambda\,\omega/\omega_p)$ in the high frequency limit.

\par 

In Silin \cite{silin65}, for the low intensity ($v_E\ll v_\mathrm{th}$) and low frequency ($\omega\ll\omega_p$) limit, it is found that $C=1$ and $\ln(\Lambda_{ei}^{(IB)})=\ln(\Lambda)$ and for the high frequency limit ($\omega\gg\omega_p$) it is found that $C=1$ and $\ln(\Lambda_{ei}^{(IB)})\approx\ln(\Lambda\,\omega/\omega_p)$ (cf. appendix \ref{sec:apA}). In the high intensity ($v_E \gg v_\mathrm{th}$) and low frequency ($\omega\ll\omega_p$) limit, it is $C=12/\sqrt{2\pi}\,(v_\mathrm{th}/v_E)^3\,\left(\ln\left(v_E/2 v_\mathrm{th}\right)+1\right)$ and $\ln(\Lambda_{ei}^{(IB)})=\ln(\Lambda)$.

\par 

In Jones-Lee \cite{jon82}, for high intensities ($v_E\gg v_\mathrm{th}$) and low frequencies ($\omega\ll\omega_p$) it is found (eq.(63) in \cite{jon82}) $C=12/\sqrt{2\pi}\,(v_\mathrm{th}/v_E)^3\,\ln\left(v_E/ v_\mathrm{th}\right)$ with $\ln(\Lambda_{ei}^{(IB)})=\ln(\Lambda\,\omega_p/\omega)$ which is at variance with Silin's results who obtained a similar value of $C$ for the low frequency limit instead of the high frequency limit.  

\par 

In Mulser \cite{mulser20}, in eq. (7.71) when $\omega\gg \omega_p$, $\ln(\Lambda_{ei}^{(IB)})=\ln(\Lambda\,\omega_p/\omega)$ with $(v_\mathrm{th}^2)^{3/2}$ in $b_\mathrm{max}$ replaced by $(v_\mathrm{th}^2+v_E^2/4)^{3/2}$ or by $(v_\mathrm{th}^2+v_E^2)^{3/2}$ in \cite{mulser01}.

\par 

In Brantov et al. \cite{brant03}, what is called the effective collision in eqs.(19) or (20) of this reference, sums up to $C=(v_\mathrm{th}^2/(v_\mathrm{th}^2+v_E^2/6))^{3/2}$ or $C=(v_\mathrm{th}^2/(v_\mathrm{th}^2+v_E^2*0.3))^{3/2}$ in our notations and $\ln(\Lambda_{ei}^{(IB)})=\ln(\Lambda\,\omega_p/\omega)$ when $\omega\gg\omega_p$ in eq.(3) of \cite{brant03}.

\bibliography{dm}

\end{document}